\documentclass[11pt]{article}  

\usepackage{tikz}
\usetikzlibrary{positioning} 
\usetikzlibrary{backgrounds}
\usepackage{tikzsymbols}
\usepackage{amsmath}
\usepackage{xcolor}
\usepackage{mathtools}
\usepackage[T1]{fontenc}    
\usepackage{booktabs}       
\usepackage{amsfonts}       
\usepackage{nicefrac}   
\usepackage{float}
\usepackage{amssymb}
\usepackage{graphicx}
\usepackage[ruled,vlined]{algorithm2e}
\usepackage{bm}
\usepackage{authblk}
\usepackage{adjustbox}
\usepackage{hyperref}

\usepackage[left=2.3cm, right=2.3cm, top=3cm]{geometry}

\usepackage[round, semicolon]{natbib} 

\usepackage{pgfgantt}
\usepackage{color}
\newcommand{\ech}{\color{black}\rm} 

\usepackage[normalem]{ulem} 

\usepackage{caption}
\usepackage{subcaption}
\usepackage{multirow}

\title{
A Predictor-Informed Multi-Subject Bayesian Approach for Dynamic Functional Connectivity 
 }

\author[1]{Jaylen Lee}
\author[2]{Sana Hussain}
\author[6]{Ryan Warnick}
\author[3]{Marina Vannucci}
\author[2]{Isaac Menchaca}
\author[4]{Aaron R. Seitz}
\author[2]{Xiaoping Hu}
\author[2,5]{Megan A. K. Peters}
\author[7]{Michele Guindani}
\affil[1]{Department of Statistics, University of California, Irvine}
\affil[2]{Department of Bioengineering, University of California Riverside}
\affil[3]{Department of Statistics, Rice University}
\affil[4]{Department of Psychology, University of California Riverside}
\affil[5]{Department of Cognitive Sciences, University of California Irvine}
\affil[6]{Microsoft Security Research}
\affil[7]{Department of Biostatistics, University of California, Los Angeles}

\begin{document}
\maketitle

\begin{abstract}
\noindent
Dynamic functional connectivity investigates how the interactions among brain regions vary over the course of an fMRI experiment. Such transitions between different individual connectivity states can be modulated by changes in underlying physiological mechanisms that drive functional network dynamics, e.g., changes in attention or cognitive effort.
In this paper,  we develop a multi-subject Bayesian framework where the estimation of dynamic functional networks is informed by time-varying exogenous physiological covariates that are  simultaneously recorded  in each subject during the fMRI experiment. More specifically, we consider a dynamic Gaussian graphical model approach where a non-homogeneous hidden Markov model is employed to classify the fMRI time series into latent neurological states.
We assume the state-transition probabilities to vary over time and across subjects as a function of the  underlying covariates, allowing  for the estimation of recurrent  connectivity patterns and the sharing of networks among the subjects. We further assume sparsity in the network structures via shrinkage priors, and achieve edge selection in the estimated graph structures by introducing a multi-comparison procedure for shrinkage-based inferences with Bayesian false discovery rate control. We evaluate the performances of our method \emph{vs} alternative approaches on synthetic data. We apply our modeling framework on a resting-state experiment  where fMRI data have been collected  concurrently with pupillometry measurements, as a proxy of cognitive processing, and assess the heterogeneity of the effects of changes in pupil dilation
on the subjects' propensity to change connectivity states. The heterogeneity of state occupancy across subjects
provides an understanding of the relationship between increased pupil
dilation and transitions toward different cognitive states.

\end{abstract}

\section{Introduction}
Functional connectivity (FC) has emerged as one of the most active research areas in functional magnetic resonance imaging (fMRI). The purpose of FC studies is to characterize the undirected statistical dependencies between brain regions  and thus gain a greater understanding of brain functioning  \citep{friston1994_1, Hutchison2013}. 
 Simple approaches to studying FC rely on readily available measures of temporal correlation, such as the partial correlations between two brain regions after conditioning upon all other regions \citep{Fornito2013, Friston2011}.  Such metrics assume that interactions between brain regions are constant in space and time throughout the fMRI session \citep[\emph{static connectivity},][]{li2008}. Rather, neuroscientists have become increasingly aware that functional connectivity is dynamic and fluctuating, i.e. non-stationary, and that studying the dynamics of  FC is crucial for improving our understanding of human brain function \citep{Hutchison2013, Vidaurre2017, Lurie2020}.  The term ``chronnectome" has been introduced to describe the growing focus on identifying time-varying, but reoccurring, patterns of coupling among brain regions \citep{Calhoun2014}. 

Recent studies have highlighted how  subjects are  more likely to experience particular connectivity states when some underlying physiological conditions are present. For example, \citet{Chand2020} have investigated the association between heart rate variations and FC. Similarly, in a sleep fMRI study, \citet{El-Baba2019} have shown that transitions between connectivity states slow as subjects fall into deeper sleep stages. As a further example, \citet{Kucyi2017} have described how connectivity dynamics are associated with attentiveness in a pencil-tapping task.  These studies, among others, have motivated the need for models that provide a better understanding of how the transitions between different functional connectivity states are modulated by internal or external conditions measured during the course of an experiment. In the experimental study we consider in this manuscript, we have available fMRI data collected together with pupillometry measurements. Pupil dilation has become increasingly popular in cognitive psychology to measure cognitive processing and resource allocation.  It is believed that the changes in pupil diameter reflect brain state fluctuations driven by neuromodulatory systems \citep{Sobczak2021}. For example, the pupil dilates more under conditions of higher attention \citep{Sigle2003}. Thus, pupil dilation measurements can be seen as an index of effort exertion, task demand, or difficulty in an fMRI experiment \citep{vanderWel2018}. Thus, it is  of interest to understand how pupil dilation is associated with an increased probability of particular connectivity states experienced by a subject during an experiment \citep{Martin2021}.  

Many of the commonly used approaches for studying dynamic connectivity  rely on multi-step inferences.  For example, in \citet{Calhoun2014} the fMRI time courses are first segmented by a sequence of sliding windows,  and then precision matrices are estimated in each window. Finally,  $k$-means clustering methods are used  to  identify re-occurring patterns of FC states. Post-hoc analyses may be employed to assess the association of the estimated dynamic connectivity states with other available measurements, like pupil dilation measurements \citep{Haimovici2017}.
However, the arbitrary choice of the window length and the offset may lead to spurious dynamic profiles and poor estimates of correlations for each brain state  \citep{Lindquist2014,SHAKIL2016111}. 
Improvements were proposed by \citet{cribben2012, Cribben2013} and \citet{Xu2015}, who developed change point detection methods to recursively partition the fMRI time series into stable contiguous segments where networks of partial correlations are estimated by employing the graphical lasso of \citep{Friedman2008}. These methods do not require pre-specifying the segment lengths and can detect the temporal persistence of the functional networks. However, they do not account for the possibility of states being revisited and hence do not conform to the idea that the chronnectome exhibits recurrent patterns of dynamic coupling between brain regions of interest (ROIs).
 
Other model-based approaches to dynamic connectivity consider the set of ROIs as the nodes (or vertices) of an underlying graph and employ homogeneous hidden Markov models (HMMs) to detect state transitions and infer a discrete set of latent connectivity states over time.\citet{Warnick2018b} develop a Bayesian HMM to model dynamic FC as the transition between state-specific graphs, each graph being related to others via an underlying super-graph. 
\citet{Sourty2016} use product HMMs to describe the evolution of the sliding-windows correlations and capture temporal dependencies across resting-state networks. \citet{Chiang2015} used a Bayesian HMM to estimate the dynamic structure of graph theoretical measures of whole-brain FC. Also, HMMs have been employed in time-varying vector autoregressive (VAR) modeling frameworks for whole-brain resting state connectivity, where the VAR coefficients and the innovation covariance matrix are allowed to change with the latent states \citep{Vidaurre2017, ting2018,ombao2018}. However, these implementations of hidden Markov models typically assume that the probabilistic model underlying the state transitions is constant throughout an experiment. Crucially, such a homogeneity assumption does not allow to assess the modulatory effect of time-varying physiological factors on the transitions, e.g. how changes in vigilance measured via pupil dilation can impact state transitions \citep{Lurie2020}.  

In this paper, we develop a multi-subject Bayesian framework where the estimation of dynamic functional networks is informed by
time-varying exogenous physiological covariates that are simultaneously recorded in each subject during the fMRI experiment. More specifically, we introduce a multi-subject non-homogeneous HMM modeling framework where the transition probabilities between states are shared between subjects and  vary over time as a fucntion of  the covariates. Our setting allows for the estimation of unique connectivity state transitions for each subject. It also permits group-based inferences, via recurring connectivity patterns and sharing of networks among the subjects. With respect to the multi-step inference strategies described above, in our approach both  the dynamic connectivity states and their association with the physiological measurements are estimated in a single modeling framework, accounting for all uncertainties.  \citet{Kundu2018} have recently proposed a  two-step multi-subject fused-lasso approach  for detecting change points in functional connectivity. Differently from their proposal, our method does not assume that the experimental design and the timing of the change points between connectivity states are shared  among all subjects, and can therefore be applied to more general experimental designs.  Indeed, our approach allows for differing state transition behavior across multiple subjects by  modeling the state transition parameters hierarchically.  
Our modeling approach further assumes sparsity in the network structures, by assuming a shrinkage prior on the connectivity networks. Additionally, we propose a strategy for edge selection that combines the posterior shrinkage-informed thresholding approach of \citet{Carvalho2010} with the Bayesian False Discovery Rate controlling method of \citet{Mueller2006}.

We apply our modeling framework to a resting-state experiment where fMRI data have been collected concurrently with pupillometry measurements, leading us to assess the heterogeneity of the effects of changes in pupil dilation on the subjects’ propensity to change connectivity states. Changes in pupil diameter have been linked to attention and cognitive efforts modulated by the activity of norepinephrine-containing neurons in the locus coeruleus (LC). For example, \citet{joshi_relationships_2016}  have shown that LC activation predicts changes in pupil diameter that either occur naturally or are caused by external events during near fixation, as in many experimental tasks. 
Therefore, pupil dilation has been used as a proxy for a metric of a person's willingness to exert more effort and involve a greater mental effort to complete a task. Recent methods for studying such association use a multi-step approach, first identifying switches in subjects’ state sequences and then calculating the difference between the normalized pupil size before and after the estimated switch \citep[see, e.g.][]{hussain_locus_2022}. 
In our application, we demonstrate how the model can recover expected change points in dynamic FC states, as those states align quite well with the experimental events regulated by the behavioral task.
\ech

The rest of the paper is organized as follows. In section 2 we describe our proposed method and edge selection procedure. We also give a brief synopsis of our Markov Chain Monte Carlo (MCMC) approach to posterior inference. In section 3 we showcase our model performance on simulated data and provide comparisons to the sliding window and homogenous HMM approaches. Lastly, in Section 4, we apply our model to the LC handgrip data with accompanying results and analysis. Section 5 concludes the paper with a discussion.

\section{Methods}
In this section, we describe our proposed predictor-informed multi-subject model for dynamic connectivity. This is a single-step fully Bayesian approach that explicitly models the heterogeneity in the dynamics of connectivity patterns across all subjects and -- simultaneously  -- estimates the predictor effects on those dynamics. We achieve this by constructing a non-homogeneous Hidden Markov Model (HMM) where the transition probabilities are functions of time-varying covariates.

\subsection{An HMM model for dynamic functional connectivity}

Let  $Y^i_t = (Y^i_{t1},\ldots, Y^i_{tR})^T$ denote the vector of fMRI BOLD responses measured at time $t$ in R regions of interest (ROIs), $t = 1,\ldots, T$ on subject $i=1, \ldots, N$. We  adopt  a Gaussian graphical model framework, and assume multivariate normality of the bold signals, that is $Y^i_t \sim N_R(\mu^i_t, \Omega_t^{-1,i})$, where $\mu^i_t$ is a mean regression term and $\Omega^i_t$ indicates a time-varying precision matrix, i.e. the inverse covariance matrix at each time point. In graphical models, the zeros of the precision matrix correspond to  conditional independence; that is, if an off-diagonal element $\omega_{jkt} = 0$, $j,k=1, \ldots, R, j\neq k$, then the signals $Y^i_{tj}$ and $Y^i_{tk} (j \neq k)$ are conditionally independent. The mean term $\mu^i_t$ can be specified as a general linear model \citep{Friston1994} to capture activation patterns over time, as done for example in \cite{Warnick2018b}. Here, however, since we are interested  in capturing connectivity patterns through the modeling of the time-varying precision matrix, we assume without loss of generality that the BOLD signal has been mean-centered by removing any estimated trend and activation component. This ``detrending'' is not uncommon for studying FC, especially for task-based fMRI data, where the data are first mean-centered, to remove any systematic task-induced variance, and the analysis is then conducted on  the time series of the residuals \citep[see, e.g.][]{FAIR2007396}.  

We propose to model the dynamics of FC using an HMM framework with $S$ latent states characterizing FC and the brain transitions during the fMRI experiment. Our formulation captures the heterogeneity of individual-specific patterns of connectivity over time, since each subject's fMRI data may be  characterized by specific change points and evolution of the brain's functional organization.  However, we assume that the connectivity patterns may also be re-occurring and they can possibly be shared among the subjects, thus allowing for group-based inferences. 
Let $ (s_1, \ldots, s_T)$ be a $T$-dimensional vector of categorical indicators $s_t$, such that $s_t=s$ if state $s$ is active at time $t$, $s=1, \ldots, S$. Then, we assume the data follow a Gaussian graphical model at time $t$ of the type
\begin{equation}\label{eq:like}
	Y^i_{t} | s^i_t = s, \Omega_{s} \sim N_R(0,\Omega^{-1, i}_{s}), \quad s=1, 
	\ldots, S,
\end{equation}
with subject-level precision matrices which, at each time, are characterized by the values of one among $S$ precision matrices, identifying which state is active at that time. Model \eqref{eq:like} therefore implies connectivity networks that vary by subjects and by state.

\subsection{Modeling connectivity transitions as a function of observed physiological factors}\label{sec:nhmm}
Many neuroscience experiments involve the simultaneous collection of fMRI data together with  physiological, kinematics and behavioral data \citep{Wilson2020}. For example, our motivating application considers a handgrip task where pupillometry dilation data (i.e., measurements of pupil dilation sizes) are concurrently recorded.  Pupillary dilation is regarded as a surrogate measure for activity in the locus coeruleus  circuit, which plays a central role in many cognitive processes involving attention and effort, and it is considered the main source of the neurotransmitter noradrenaline, a chemical released in response to pain or stress. Neuronal loss in the locus coeruleus is known to occur in neurodegenerative disorders such as Alzheimer's disease and related dementias as well as Parkinson's disease dementia. It is therefore important to understand how brain dynamics may be differentially modulated as a function of pupil dilation in different subjects. 

Here, we propose to model the dynamics of FC by developing a non-homogeneous HMM framework where estimation is informed by subject-level time-varying exogenous physiological covariates, e.g. physiological factors like the pupillary data in our motivating application.  More in detail, we assume that switches between states are regulated by transition probabilities that vary over time and across subjects as a function of $B$ time-varying subject-level covariates as
\begin{equation}
Q^i_{rst} =P(s_{t+1}=s \, | s_t=r)= \frac{\exp(\xi^i_{rs}+\bm{x}_t^{i^T} \bm{\rho}^i_s)}{\sum_{l = 1}^S \exp(\xi^i_{rl}+\bm{x}_t^{i^T} \bm{\rho}^i_l)}, \quad r,s=1, \ldots, S, 
\label{eq:trans_model}
\end{equation}
where  $\bm{x}_t^{i}$ denotes a $B\times 1$ vector of covariate values for subject $i$ at time $t$, and $\bm\rho^i_s=(\rho^i_{s1}, \ldots,\rho^i_{sB})$ is the corresponding $B\times 1$ vector encoding the effect of each covariate on the probability of transitioning to state $s$ for subject $i$. The parameter $\xi^i_{rs}$ defines a baseline transition probability from state $r$ to state $s$ for subject $i$, that is the transition probability without any covariate effect. To ensure identifiability,  we define a  state  as reference. Without loss of generality, we set $s=1$ as the reference state, and also set the coefficients $\rho^i_{1b}$,  $b=1, \ldots, B$, and $\xi^i_{1\, \cdot}$, $i=1, \ldots N$ equal to zero. Thus, the state transition coefficients are interpreted with respect to the reference state, and we can re-express  \eqref{eq:trans_model}  in terms of the log-relative odds of the transition from state $r$ to state $s$ compared to the  transition from state $r$ to the reference state 1, 
\begin{equation}
 	log(\frac{Q^i_{rst}}{Q^i_{r1t}}) = \xi^i_{rs} + \bm{x}^{i^T}_t \bm\rho^i_s, \quad r,s=1, \ldots, S.
 	\label{eq:logrelat}
 \end{equation}
In this formulation, the transition coefficients $\exp(\rho^i_{sb})$, $b = 1, \ldots B$, are more naturally interpreted as the relative change in odds of transitioning to state $s$ compared to transitioning to state $1$, after a one unit change in $x^i_{tb}$, holding all other covariates as constant.  Similarly, the coefficient $\exp(\xi^i_{rs})$ is interpreted  as  the expected odds of transitioning from state $r$ to $s$ compared to transitioning from state $r$ to $1$, when the time-varying covariates, $\bm x^i_t$, are 0 or  at a baseline/average value.
 
We assume independent Gaussian priors for  the transition parameters $\rho$ and $\xi$. We further allow for sharing of information in estimating the state transition structure across subjects, by  employing a hierarchical modeling formulation for the state transition parameters. More specifically, we  assume that the individual coefficients $\xi_{rs}^i$ and $\rho_{sb}^i$, $b=1,\ldots, B$, vary around population-level means, $Z_{rs}$ and $\eta_{sb}$,  as follows:  
 \begin{align} 
 \begin{split}
 	s^i_{t+1} | s^i_t &= r \sim Multi(Q^i_{r,\cdot,t} ) \quad t=1, \ldots, T, \\ 
 	\xi^i_{rs} &\sim N(Z_{rs}, \sigma_\xi ),\\ 
 	\rho^i_{sb} &\sim N(\eta_{sb}, \sigma_\rho ), \\ 
 	Z_{rs} &\sim N(z^0_{rs}, \sigma_z ), \\ 
 	\eta_{sb} &\sim N(0, \sigma_\eta ),
 	\label{eq:hier_model_states}
 	\end{split}
 \end{align}
where $Q^i_{r,\cdot,t}=(Q^i_{r,1,t}, \ldots, Q^i_{r,S,t})^T$, and $r,s=1, \ldots, S$, $b=1, \ldots, B$. By allowing each subject to have their own transition parameters the model allows for unique subject-level transition behavior while also capturing population-level estimates through the group level parameters.  The interpretation of the group level parameters, $\eta$ and $Z$, is similar to their single subject counterparts. The prior means $z^0_{rs}$ are usually set to 0 except for $z^0_{rr}$, $r \neq 1$, which is set to be positive to encourage state persistence over time (self-transitions) and thus more stable estimated state sequences. Keeping in mind that these state transition parameters operate on the log odds of transition relative to state 1, and that interpretation of the parameters require exponentiation, a small shift in value for the state transition parameters can result in rather large changes in state transition behavior. To this end, we recommend setting the variance parameters of the priors for $\xi$, $\rho$, $Z_{rs}$ and $\eta_{sb}$ to some small positive value on the order of $0.1$.

\subsection{Modeling sparsity through a graphical horseshoe prior}\label{sec:GraphicalHorshoe}
\begin{figure}[t!]
    \centering
    \resizebox{0.45\columnwidth}{!}{%
    \begin{tikzpicture}[
roundnode/.style={circle, draw=gray, very thick, minimum size=8mm},
squarednode/.style={rectangle, draw=gray, very thick, minimum size=8mm},
blank/.style={rectangle, draw=white, very thick, minimum size=8mm},
node distance=1cm, framed
]

\node[roundnode] (st1) {$s^i_{t}$};
\node[roundnode] (st2) [right=of st1] {$s^i_{t+1}$};
\node[roundnode] (rho) [above =of st1] {$\rho^i_{s_t}$};
\node[roundnode] (xi) [above =of st2] {$\xi^i_{s_t,\cdot}$};
\node[roundnode] (z) [above=of xi] {$Z_{s_t,\cdot}$};
\node[roundnode] (Eta) [above=of rho] {$\eta_{s_t}$};
\node[squarednode] (sigma_eta) [left=of Eta] {$\sigma_\eta$};
\node[squarednode] (sigma_z) [right=of z] {$\sigma_\rho$};
\node[squarednode] (sigma_rho) [left=of rho] {$\sigma_\rho$};
\node[squarednode] (sigma_xi) [right=of xi] {$\sigma_\xi$};
\node[roundnode] (yt1) [below=of st1] {$Y^i_{t}$};
\node[roundnode] (yt2) [below=of st2] {$Y^i_{t+1}$};
\node[ below = 1cm of yt1](id1){$s^i_t\in\{1, \ldots, S\}, t=1, \ldots, T$, $i=1, \ldots, N$};

\draw [->] (st1) to [out=330,in=210] (st2);
\draw[->] (xi.south) -- node[anchor=east] {} (st2.north);
\draw[->] (rho.south) -- node[anchor=north] {$x^i_{t}$} (st2.west);
\draw[->] (st1.south) -- node[anchor=east] {$\Omega_{s^i_{t}}$} (yt1.north);
\draw[->] (st2.south) -- node[anchor=east] {$\Omega_{s^i_{t+1}}$} (yt2.north);
\draw[->] (z.south) to (xi.north);
\draw[->] (Eta.south) to (rho.north);
\draw[->] (sigma_rho.east) to (rho.west);
\draw[->] (sigma_xi.west) to (xi.east);
\draw[->] (sigma_eta.east) to (Eta.west);
\draw[->] (sigma_z.west) to (z.east);

\end{tikzpicture}
}
    \caption{Graphical representation of the proposed PIBDFC. The data $Y^i_t$ are emissions from a distribution that is parameterized by a precision matrix $\Omega_{s_t^i}$, which encodes the FC and is determined by the state active at time $t$: $s^i_t\in\{1, \ldots, S$, $t=1, \ldots, T$, $i=1, \ldots, N$. The probabilities of transitions from $s^i_t$ to $s^i_{t+1}$ are given by the $(s^i_t, s^i_{t+1})$ entry of the $S\times S$ matrix $Q^i_{\cdot,\cdot, t}$. This entry is modeled according to Equation \ref{eq:logrelat}.}
    \label{fig:graph_model}
\end{figure}
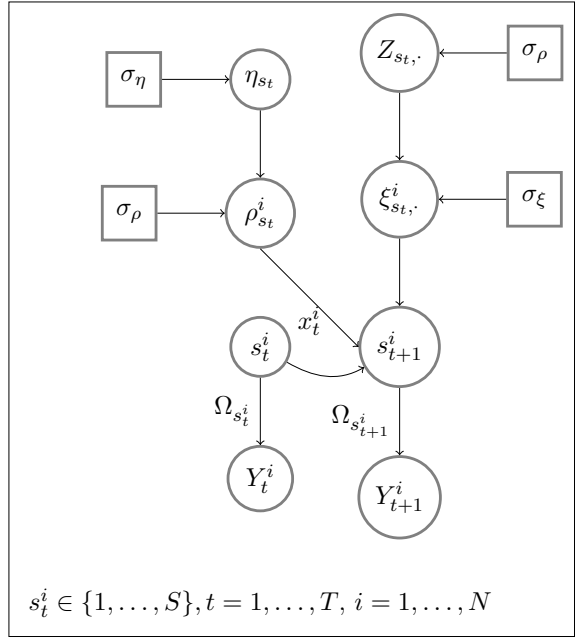

Functional networks are thought to exhibit the so-called small world behavior, indicating a high degree of clustering and high efficiency in the estimated networks \citep{Wang2010, Essen2016}. This leads to an expectation of sparsity within functional networks and the associated precision matrices. In a Bayesian framework, sparsity can be enforced by postulating either selection- or shrinkage-inducing priors. Selection involves  inferring which off-diagonal elements of the precision matrix should be set to exact zeros. \citet{Warnick2018b} achieve such a selection by using a G-Wishart prior to sample positive definite matrices  according to  estimated adjacency matrices that correspond to the FC networks. This selection approach is intuitive and leads to straightforward inferences via the posterior probabilities of inclusion of the elements of the precision matrix. However, it is computationally challenging and does not scale well to relatively large dimensions of the networks. Here, instead, we take a shrinkage-based approach and model the off-diagonal entries of the state-specific precision matrices $\Omega_s$, $s=1, \ldots,$ in \eqref{eq:like} by employing a graphical horseshoe prior \citep{Li2019b}.  Thus, we set
\begin{equation}
p(\Omega_s | \tau, \Lambda )\propto \prod_{j<k} N(\omega_{jks} | \lambda_{jk}^2\, \tau^2)\prod_{j<k} C_+ (\lambda_{jk} | 0,1)I(\Omega_s \in S_R), \quad s=1, \ldots, S
\label{eq:precision_prior}
\end{equation}
where $I(\Omega_s \in S_R)$ is an indicator function to ensure that samples of $\Omega_s$ belong to the space of positive definite $R\times R$ matrices and $C_+(\cdot; \mu, \sigma)$ denotes a  half-Cauchy distribution with  location parameter $\mu$ and scale $\sigma$. In \eqref{eq:precision_prior}, we further assume  a non-informative flat prior  for the diagonal elements, i.e. $\omega_{jjt}\propto 1$. The shrinkage of the off-diagonal elements is obtained through the combined effect of the variance components $\lambda_{jk}^2$ and $\tau^2$ in the normal priors for $\omega_{jkt}$, $j=1, \ldots, k-1$, $k=1, \ldots, R$. The parameter $\tau$ is a \textit{global shrinkage parameter}, that controls how sparse the precision matrix is in its entirety. The parameter $\lambda_{jk:j<k}$ defines instead a \textit{local shrinkage parameter}, since it allows to  shrink each individual off-diagonal entry $\omega_{jk}$ towards zero, whereas it maintains the magnitude of non-zero entries and thus reduces biases. Following \citet{Li2019b}, we assume a half-Cauchy prior on $\tau$, $\tau  \sim C_+(\cdot; 0,\tau_0)$, with $\tau_0$ indicating an \textit{a priori} belief about the global sparsity of the estimated graph. In order to specify $\tau_0$, one can simulate graphs under the informal selection rule of \citet{Carvalho2010}, where an edge j,k is selected if $E(\frac{1}{1+\lambda_{jk}\tau}) < 0.5$. We demonstrate such a process in Figure \ref{tau_0_effects} in the Appendix. We find that a \(\tau_{0}=1\) gives an expected edge density of approximately \(50 \%\) while having the
largest spread. Figure \ref{fig:graph_model} provides a graphical representation of the proposed predictor-informed Bayesian  dynamic FC model (PIBDFC).

\subsection{Posterior Inference}

The posterior distribution for the parameters in the proposed model is not available in closed form. Hence, we revert to Markov Chain Monte Carlo (MCMC) techniques for posterior inferences. In particular, we follow \citet{Holsclaw2017} and employ Polya Gamma auxiliary variables \citep{Polson2013} to sample the state transition parameters. Based on the sampled $Q^i_{\cdot, \cdot, t}$, we can construct a sequence of transition matrices based on equation \eqref{eq:logrelat}. After normalizing each row $Q^i_{s, \cdot, t}$ so that it sums to 1, we use a stochastic forward-backward algorithm  to sample the state sequence \citep{Scott2002}. Then, conditioned upon the estimated state sequence, it is possible to sample the precision matrix parameters using the blocked Gibbs algorithm presented in \citet{Li2019b}. Other parameters in the hierarchical model for the states' transitions \eqref{eq:hier_model_states} are sampled via simple Gibbs steps. By iterating through the steps above, we obtain samples from the posterior. We provide a brief summary below:

\begin{itemize}

 \item[1.] \textbf{Sample $\mathbf{ Q^i_{\cdots}, \xi^i_{\cdot \cdot}, \rho^i_\cdot }$:} 
We can rewrite the likelihood for $\xi^i_{rs}$ according to \citet{Holmes2006} to be in the form of Equation \ref{holmeseq}.
\begin{equation}
   L(\xi^i_{rs}) \propto \prod_{t: s^i_{t-1} = r} \frac{exp(\xi^i_{rs}-c^i_{rst})^{I(s^i_t = s)}}{1+exp(\xi^i_{rs}-c^i_{rst})}
   \label{holmeseq}
\end{equation}

where $c^i_{rst} = log\sum_{m \neq s} exp(\xi^i_{rm}+\mathbf{x^i_t \rho^i_m} - \mathbf{x^i_t \rho^i_s})$. Using the Polya-Gamma augmented logistic regression technique of \citet{Polson2013}, we get the posterior of $\xi^i_{rs}$ to be conditionally Gaussian. 

$$\xi^i_{rs} | \cdot \sim N\left(\frac{Z_{rs}/\sigma_\xi + n_{rsi} - N_{ri}+2\sum_{t:s^i_{t-1} = r}\omega^i_{rst}c^i_{rst}}{\sum_{t:s^i_{t-1} = r}\omega^i_{rst}+1/\sigma_\xi},  (\sum_{t:s^i_{t-1} = r}\omega^i_{rst}+1/\sigma_\xi)^{-1}\right)$$

where $n_{rsi}$ is the count of transitions from state $r$ to state $s$ during the timecourse of subject $i$ and $N_{ri}$ is the number of times subject $i$ visited state $r$. $\omega^i_{rst}$ is a Polya-Gamma random variable distributed $PG(1, \xi^i_{rs} - c^i_{rst})$. We use a similar strategy to update $\rho^i_{rb}$, the logistic component for subject $i$ for state $r$ and covariate $b$, achieving the posterior:

$$\rho^i_{rb} | \cdot \sim N\left(\frac{\eta_{rb}/\sigma_\rho + \sum_{t = 1}^{T_i}x^i_{tb}(I(s_{t+1} = r)-1/2+\omega^i_{rbt}c^i_{s_trb})}{\sum_{t = 1}^{T_i}(x^i_{tb})^2\omega^i_{rbt}+1/\sigma_\rho},  (\sum_{t = 1}^{T_i}(x^i_{tb})^2\omega^i_{rbt}+1/\sigma_\rho)^{-1}\right)$$

where $c^i_{rst} = log\sum_{m \neq s} exp(\xi^i_{rm}+\mathbf{x^i_t \rho^i_m} - \mathbf{x^i_t \rho^i_s})$.

\item[2.]  \textbf{Sample $s^i_t$:} 
We sample the sequence of states by adapting the stochastic forward-backward algorithm presented by \citep{Scott2002}.

\item[3.] \textbf{Sample the matrices $\Omega_{s}^i$, $s=1, \ldots, S$}: The conditional posterior for $\Omega_s$ is as follows:

$$P(\Omega_s| \mathbf{Y}, s^\cdot_\cdot, \lambda_{\cdot \cdot s}, \tau_s) \propto \prod_{\{i,t: s^i_t = s\}} N_R(Y^i_t| 0, \Omega_s^{-1}) \prod_{j = 2}^R \prod_{i=1}^j N(\omega_{ijs}| 0, \lambda_{ijs}\tau_s)$$

For MCMC inference purposes, \citet{Li2019b} adopt auxiliary variables $\nu_\lambda$ and $\xi_\tau$, in order to modify the Gibbs sampling procedure presented by \citet{Makalic2016}. This procedure is performed for a column-wise update in a fashion similar to \citet{Wang2012}. For each state, we update $\Omega_s$ by following the Graphical Horseshoe algorithm letting $S = n_s*\hat{\Sigma_s}$ where $n_s$ and $\hat{\Sigma_s}$ are the sizes and sample covariance matrices of observations assigned to state $s$. 

\item[4.] \textbf{Sample $\mathbf{Z_{rs}, \eta_b}$:} These conditional posteriors follow the typical normal-normal update:
$$Z_{rs} | \cdot \sim N\left((\frac{1}{\sigma_z}+\frac{n}{\sigma_\xi})^{-1}
(\frac{z^0_{rs}}{\sigma_z}+\frac{\sum_i \xi^i_{rs}}{\sigma_\xi}),(\frac{1}{\sigma_z}+\frac{n}{\sigma_\xi})^{-1}\right),$$

$$\mathbf{\eta_{b}} | \cdot \sim N\left((\frac{1}{\sigma_\eta}+\frac{n}{\sigma_\rho})^{-1}
(\frac{\mathbf{\eta^0_{b}}}{\sigma_\eta}+\frac{\sum_i \mathbf{\rho^i_{b}}}{\sigma_\rho}),(\frac{1}{\sigma_\eta}+\frac{n}{\sigma_\rho})^{-1}\right).$$

\end{itemize}

\subsection{Graph Selection}
\label{sec:graph_sel}
Our model achieves sparsity of the estimated functional network thanks to the shrinkage of the off-diagonal elements of $\Omega$ provided by the graphical horseshoe prior. However, shrinkage priors do not lead to exact zeros. Hence, non-relevant connectivities need to be identified through post-MCMC inference. For example,  \citet{Li2019b} suggest using 50\% posterior credible intervals of the inverse-covariance elements, and then thresholding the off-diagonal element to zero if the interval contains 0, reporting the posterior mean otherwise. However, the resulting selection does not provide a multiplicity control for false discoveries. 

We follow a decision-theoretic approach and formulate the graph selection problem as a testing problem based on the posterior evidence of shrinkage for each off-diagonal element of the precision matrix $\Omega_s$. Since we consider the posterior estimates of $\Omega_s$ for each state $s=1, \ldots, S$, separately,  in the following we drop the superscript $s$ for notational simplicity, unless needed for clarity. For any given state $s=1, \ldots, S$, the $j,k$ off-diagonal element $\omega_{jk}$ $(j<k; k = 2,\ldots,R)$ provides a measure of the connectivity level, with $\omega_{jk}=0$ indicating that the connectivity is truly zero, and $|\omega_{jk}|\neq 0$ otherwise.  Let $\delta_{jk}$ indicate the decision (action) in the testing problem, that is $\delta_{jk}=1$ corresponds to rejecting the null hypothesis of no connectivity and  $\delta_{jk}=0$ failure to reject (acceptance). Let $D=\sum_{j<k}\delta_{jk}$ indicate the total number of positive (significant) decisions taken. Following \citet[see][]{Muller07}, for  real numbers $c_1, c_2 >0$, we can then determine the optimal set of decisions $\bm\delta=\{\delta_{12}, \delta_{13}, \ldots, \delta_{R-1\, R}\}$ by minimizing the following loss function:
\begin{equation*}
L_{\Omega_s}(\Omega_s, \bm\delta, \bm Y)=-\sum_{j<k} \delta_{jk}\,  |\omega_{jk}|+c_1 \,  \sum_{j<k}\left(1-\delta_{jk}\right)\,  |\omega_{jk}|+c_2 D.
\label{eq:loss_function}
\end{equation*}
The loss function compounds a reward for correct decisions (true positives), provided by the first addend, $-\sum_{j<k} \delta_{jk}\,  |\omega_{jk}|$, where each correct decision is proportional to \(|\omega_{jk}|\)'s, and a penalty for false negative discoveries, represented by the second addend,
$\sum\left(1-\delta_{jk}\right)\,  |\omega_{jk}|$. 
The last term encourages parsimony, by increasing the penalty as the number of significant elements increases.  The optimal decision is obtained by minimizing the posterior expected loss,
\begin{equation*}
E(L_{\Omega_s}|\bm{Y}, \tau )= -\sum_{j<k} \delta_{jk}\,  E(|\omega_{jk}||\bm{Y}, \tau ) +c_1 \,  \sum\left(1-\delta_{jk}\right)\,   E(|\omega_{jk}||\bm{Y},\tau ) +c_2 D,
\end{equation*}
where $E(\omega_{jk}|\bm{Y},\tau)$ is the posterior mean of the off-diagonal elements of the inverse matrix $\Omega$.  The minimizer corresponds to a threshold of the posterior means to identify the non-zero elements of the precision matrix,
$$
\delta_{jk}^{*}=I\left\{ E(|\omega_{jk}||\bm{Y})\geq c_2 /(1+c_1)\right\}.
$$ 
\citet{Li2019b} show that the posterior mean is unbiased and it can be represented as a linear function of a shrinkage factor defined by the expected value of the random variable $\kappa_{jk}=\frac{1}{1+\lambda_{jk}^2\tau^2}$, which has a compound confluent hypergeometric distribution \citep{Gordy1998}. More in detail, $E(\omega_{jk}|\bm{Y}, \tau)=\left(1-\mathrm{E}\left(\kappa_{j k}| \bm{Y}, \tau\right)\right) \hat{\omega}_{jk}^{\prime}$ with $\hat{\omega}_{jk}^{\prime}$ representing the least square estimate of $\omega_{jk}$, $j<k$.
See Theorem 4.1 in \citet{Li2019b}, 
and related discussions in \citet{Bhadra2019}.  The result extends trivially to the folded normal distribution characterizing $|\omega_{jk}|$. Note that $\kappa_{jk}\in (0,1)$, and that larger values of $E(\kappa_{jk})$ indicate stronger shrinkage of the posterior estimates toward zero.

Graph selection can be conducted by thresholding an estimate $\hat{\kappa}_{jk}$ of the shrinkage factor $\kappa_{jk}$, i.e.
$$
\hat\delta_{i}^{*}=I\left\{ \hat\kappa_{jk}\leq \eta\right\},
$$
for some threshold $\eta\in(0,1)$. For example, in the simple regression case,  \citet{Carvalho2010} have previously recommended an informal  decision rule thresholding $\omega_{jk}$ to 0 if $1-\hat{\kappa}_{jk} < 0.5$ where $\hat\kappa_{jk}$ is the posterior median of $\kappa_{jk}$. However, such a rule does not take into account the multiplicity problem induced by the selection of the off-diagonal elements of the precision matrix. The posterior medians  $\hat\kappa_{jk}$ provide a measure of the evidence in favor of the null hypothesis, $H_0: \omega_{jk}=0$. 
Hence, a threshold $\eta$  could be set by controlling a measure of the Bayesian False discovery rate  \citep[BFDR,][]{Newton04} at a certain level $q^*$, that is to satisfy the equation
$$
BFDR(\eta) = \frac{\sum_{jk} \hat{\kappa}_{jk}\, I(\hat{\kappa}_{jk} \le \eta)}{\sum_{jk} I(\hat{\kappa}_{jk} \le \eta)} < q^*.
$$
For a related but different solution to the problem of graph selection, see also \citet{Chandra2021}, who consider inference on the partial correlation matrix derived from $\Omega$.   

\begin{figure}[t!]
	\centering
	\begin{adjustbox}{scale=0.85}
	\centering
			\includegraphics[width = 0.95\textwidth, clip]{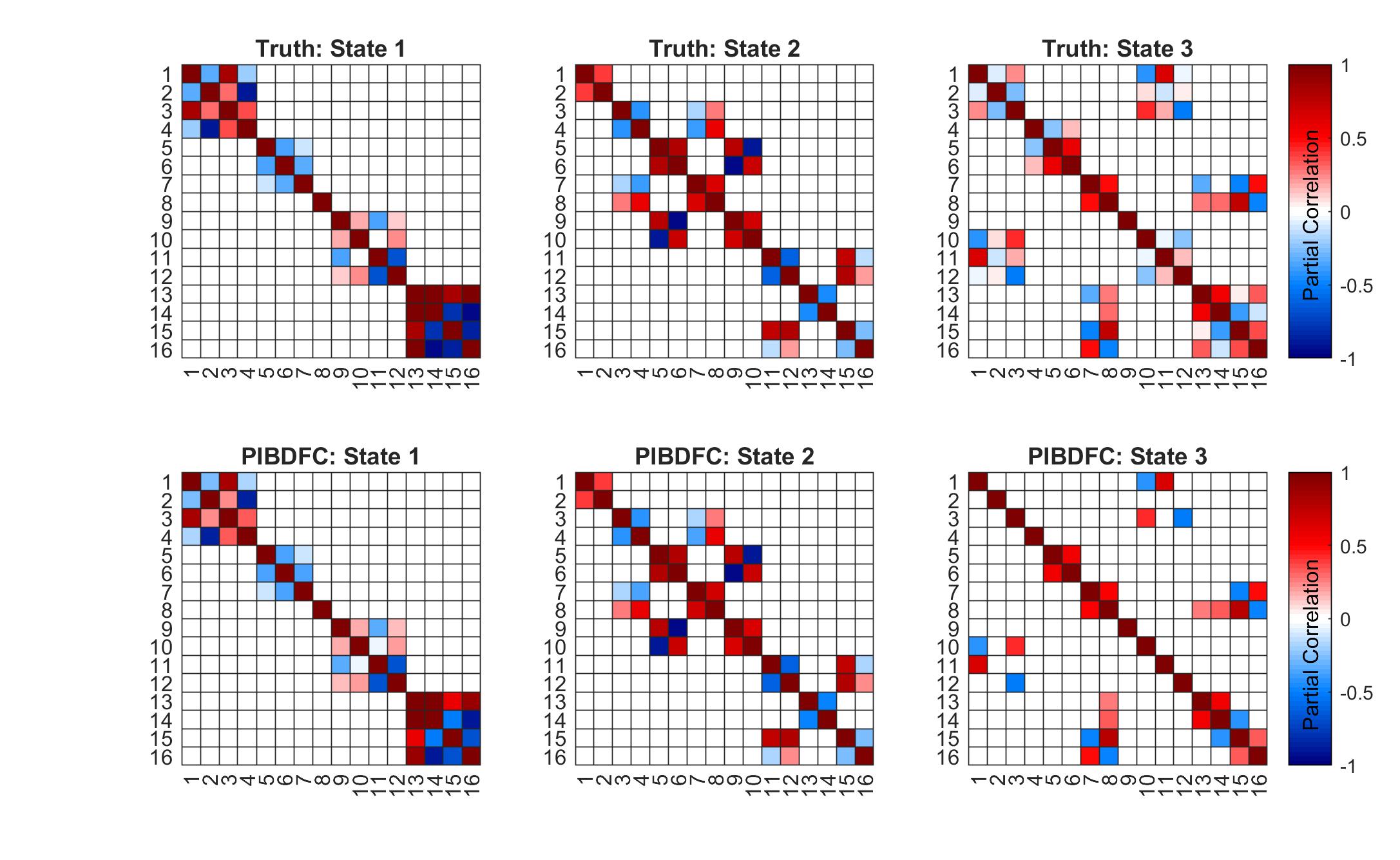}
	\end{adjustbox}
	\caption{Simulation Study 1: \textit{Top:} The true partial correlation matrices for each state responsible for generating the simulation data in the Simulation Study 1. \textit{Bottom:} The estimated partial correlation matrix from the proposed PIBDFC from a single repetition of the simulation. Each estimated partial correlation is the posterior mean of their respective distributions. Cells are set to 0 in post-hoc MCMC by controlling the BFDR at the 0.2 level. See Sections \ref{sec:graph_sel} and \ref{sec:simulation} for details.}
	\label{sim_conn_mat}
\end{figure}

\section{Simulation Study}
\label{sec:simulation}
In this Section, we present three sets of simulated datasets that aim at measuring the performance of our model with respect to the detection of non-zero connectivities and the  estimation of the latent connectivity states over time. 
More specifically, in the first two simulation studies, we compare the proposed predictor-informed Bayesian dynamic functional connectivity (PIBDFC) model with two alternative models: a widely-used tapered sliding window (Tapered SW) approach, first outlined by \citet{Allen2014}, and the Bayesian Dynamic Functional Connectivity (BDFC) model developed by \citet{Warnick2018b}. The Tapered SW represents a standard approach in the FC literature, whereas BDFC uses a homogeneous HMM to model latent connectivity state dynamics. The BDFC provides a model-based estimation of exact zeros in the functional networks at the cost of computational scalability and speed, as opposed to our computationally faster soft-shrinkage-based approach. Furthermore, the BDFC does not incorporate any predictor information in the latent state dynamics. 
Both competing approaches were developed for single-subject inference. 
We compare to our multi-subject model by concatenating the multi-subject data along the time axis for input into the respective algorithms.  All models are run on a Linux computer with an Intel Xeon Gold processor (2x 3.10 GHz) and 4 GB of RAM. For both the PIBDFC and BDFC, we simulated 5,000 posterior samples after 5,000 burn-in draws.  When fitting PIBDFC, we set the hyperparameters $\tau_0 = 1$, $\sigma_\xi=\sigma_\rho = \sigma_z = \sigma_\eta= 0.1$, following the motivations of Section \ref{sec:nhmm}. 

We assess the performance of our model on states' reconstruction by computing a set of metrics for each latent state separately. Let $r_{jk}$, $j<k; k = 2,\ldots, R$, denote the binary indicator of a non-zero connection between regions $j$ and $k$.  Following the discussion in Section \ref{sec:graph_sel}, let ${\delta_{jk}}$ indicate the decision after the model fit. Then we define the \textit{edge true positive rate} (TPR) as ${\sum r_{jk} \delta_{jk}}/{\sum r_{jk}}$. Similarly, the \textit{edge true negative rate} (TNR) is defined as ${\sum \left( 1 - r_{jk} \right) \left( 1- \delta_{jk} \right) }/{\sum \left( 1 - r_{jk} \right)}$. The \textit{Edge F1 score} (F1) is the product of the TNR and TPR, and serves as a measure of the overall  performance in graph estimation, balancing between the TPR and TNR. Analogously, we define a metric to assess the performance of the model in the estimation of the states' sequences. Let $s_t^i$ indicate the true latent state active at time $t$ for subject $i$ and let $\hat{s}_t^i$ indicate its model estimate. Then, the \textit{state sequence accuracy} for state $s$ is defined as ${\sum \{ I(s_t^i = s) I(\hat s_t^i = s)\}}/{\sum I(s_t^i = s) }$.

\begin{figure}[t!]
	\begin{center}
		\begin{adjustbox}{scale=0.75}
			\centering
			\includegraphics[width = 0.95\textwidth, clip]{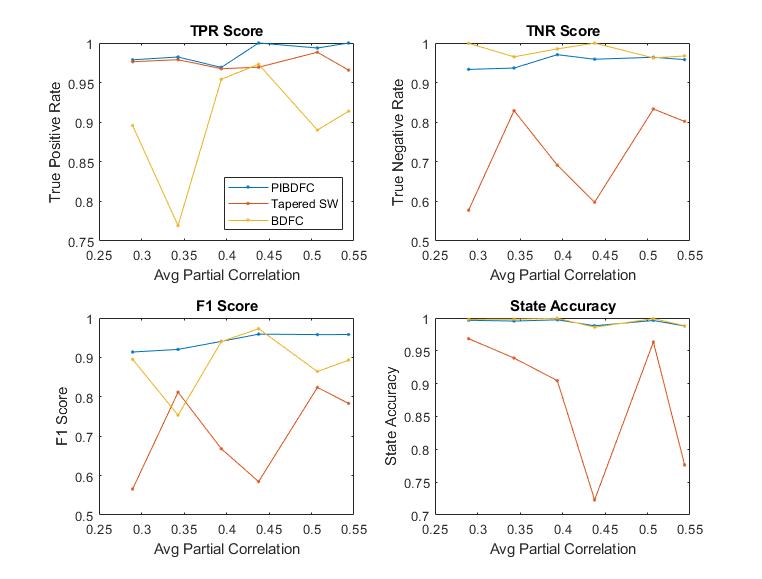}
		\end{adjustbox}
	\end{center}
	\caption{Simulation Study 1: True Positive Rate, True Negative Rate, F1 Score, and state accuracy metrics for the PIBDFC, BDFC, and Tapered SW approaches over different settings of the correlation structure. Along each horizontal axis is the average strength of the non-zero partial correlations for each state, corresponding to different levels of signal strength.}
	\label{sim1_fig}
\end{figure}

\paragraph{Simulation Study 1:}
In our first study, we investigate the performance of our model in an ideal setting where the data generation process matches the model closely. We set $T=300$ time points, $R=16$ ROIs, $N = 30$ subjects, and $S=3$ connectivity states. In this setting, we simulate data $Y^{i}_t \sim N_{16}(0, \Omega_{s^i_t}^{-1})$ with $\Omega_{s^i_t}$ encoding the individual  conditional independence structure at time $t$, identified by the value of the state indicator variables $s^i_t\in \{1, 2, 3\}$ and the prespecified graphs  in the first row of Figure \ref{sim_conn_mat}.  
In order to study the effect of the predictor information on the estimation of the transition probabilities and the FC dynamics, we introduce a single binary time-varying predictor variable, $x_t$, which transitions from $0$ to $1$ when $t = \frac{T}{2}$. For each value of the exogenous variable, we set the transition probabilities for the latent state trajectories as follows
\begin{center}
$Q_t=$
$\begin{bmatrix}
0.98 & 0.02 & 0\\
0.1 & 0.9 & 0\\
0 & 0.5 & 0.5
\end{bmatrix}$
when $x_t = 0$
;
$Q_t=$
$\begin{bmatrix}
0 & 0.5 & 0.5\\
0 & 0.7 & 0.3\\
0 & 0.02 & 0.98
\end{bmatrix}$
when $x_t = 1$.
\end{center}
Therefore, for each subject, the state sequence enforces transitions between states 1 and 2 for the first half of the time series, 
whereas it enforces transitions between states 2 and 3 for the second half.  We then simulate different state sequences for each subject using equation \eqref{eq:logrelat}, and replicated the process over 30 independent simulated data sets. In order to assess the performance of the methods for different levels of signal strength, we repeated the simulation experiment using different precision matrices $\Omega_s, s = 1,2,3$ of the same structure of the top row of Figure \ref{sim_conn_mat} but allowing for different values of the non-zero entries. This is done by using the \textit{sprandsym} function from the Mathematics toolbox of Matlab. This function takes in an adjacency matrix representation of a graph, $A_s \in \mathcal{R}^{R \times R}$ where $A_{ijs} = I(\omega_{ijs} \neq 0)$, and outputs a positive definite matrix with the same placement of 0's but random non-zero entries. This output matrix is then normalized to a partial correlation matrix. Thus, we obtained a total of six sets of precision matrices to learn the structure of. We show the aggregated results in Figure \ref{sim1_fig}. The horizontal axis reports the average strength of the non-zero partial correlations for each of the six sets of precision matrices, 
indicating a level of signal strength. The PIBDFC consistently performs better in connectivity estimation with regard to true positive rate and F1 score, across all levels of partial correlations. The BDFC appears as the most conservative, as highlighted by the large true negative rates, but low true positive rates.  Based on the results above, the PIBDFC displays the best balance of finding true non-zero partial correlations while controlling for false positives.   

\begin{figure}[t!]
	\begin{center}
		\begin{adjustbox}{scale=0.75}
			\includegraphics[width = 6.5in,height=4.5in, clip]{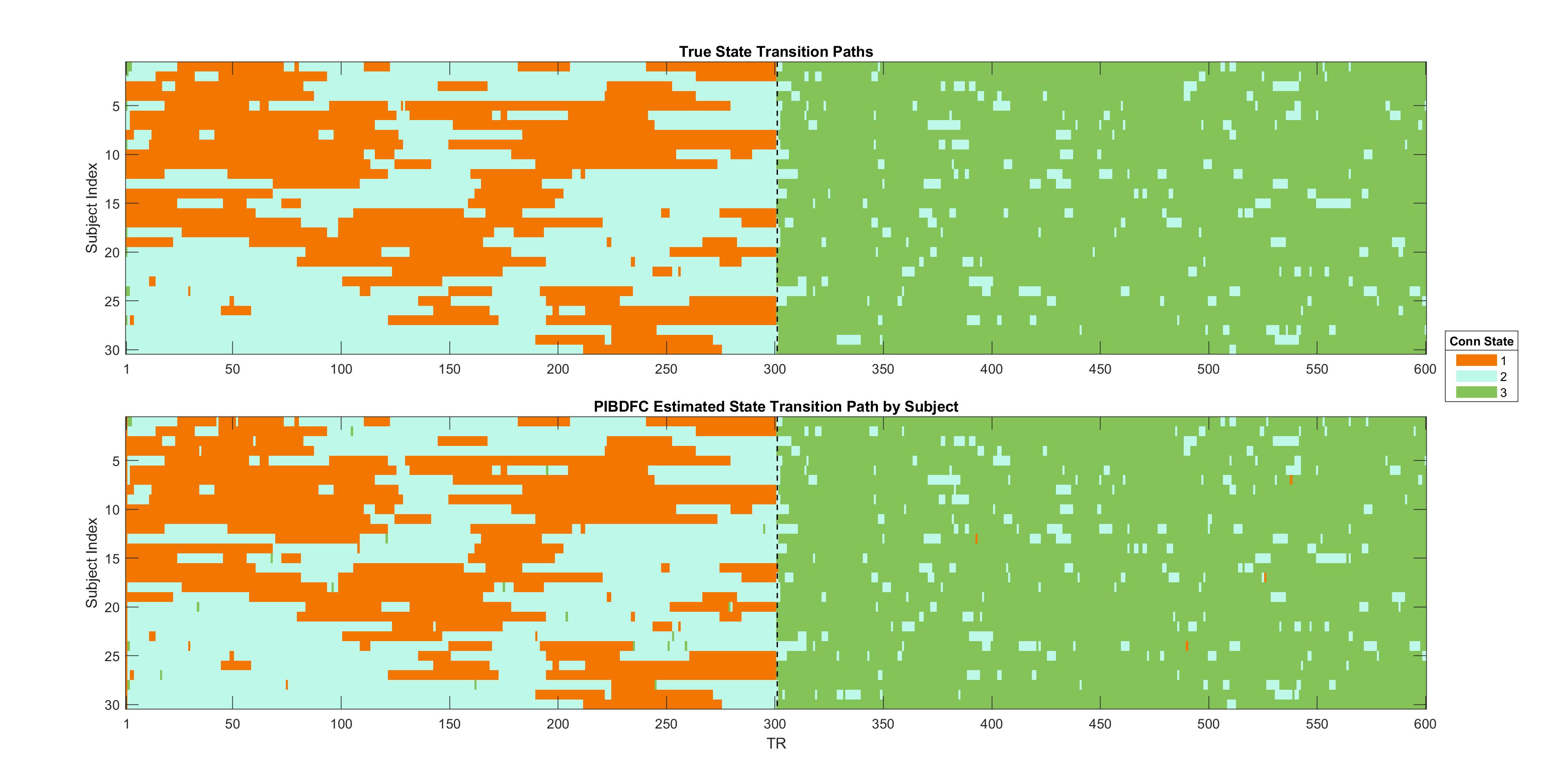}
		\end{adjustbox}
	\end{center}
	\caption{Simulation Study 1: \textit{Top:} The true state transition path for each subject (vertical axis) across each time point (horizontal axis). The color in each cell identifies which precision matrix in Figure \ref{sim_conn_mat} generated the simulated the data for each subject-time point pairs. \textit{Bottom:} The {\it maximum a posteriori} estimated state trajectories from PIBDFC. }
	\label{sim_state_mat}
	
\end{figure}

\begin{figure}[t!]
	\begin{center}
	\begin{adjustbox}{scale=0.75}
		\centering
		\includegraphics[width = 0.95\textwidth, clip]{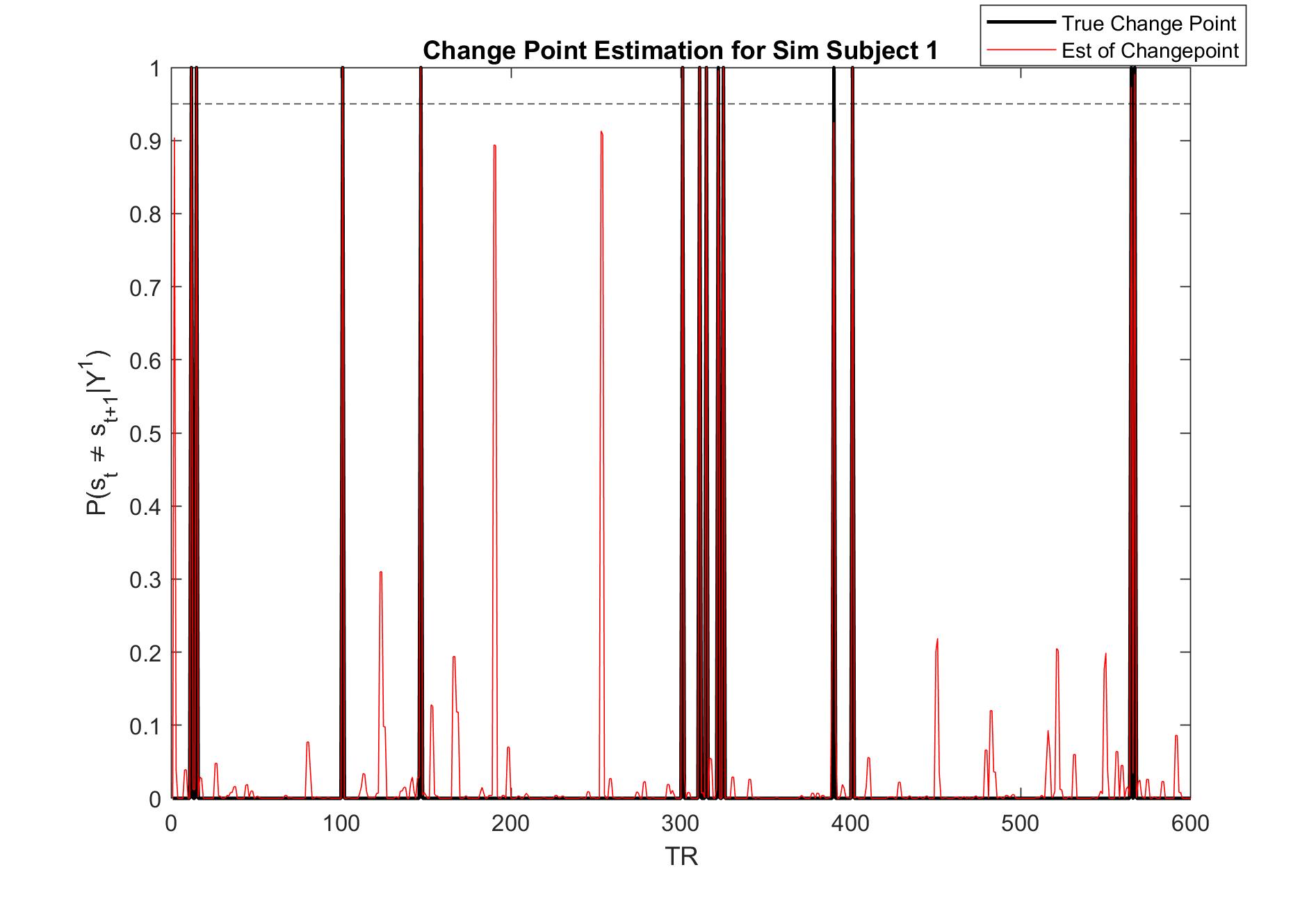}
	\end{adjustbox}
\end{center}
	\caption{Simulation Study 1: Estimation of the connectivity change points in a representative subject. The horizontal axis indicates the time points while the vertical axis reports the  posterior probability $P(s_t^1 \neq s_{t-1}^1 | Y^i_{1:T})$. The posterior probabilities of a change point are in red,  whereas the black spikes represent the true change points for the subject. We also display a horizontal dotted line at 0.95 to reflect the informal rule of declaring a change-point if $P(s_t^1 \neq s_{t-1}^1 | Y^i_{1:T}) > 0.95$.}
	\label{sim_change_point}
\end{figure}

In the following, we illustrate the inferential analyses enabled by the proposed  PIBDFC approach by showcasing a single replicate. In Figure \ref{sim_conn_mat} (bottom row) we show how the PIBDFC is able to recover the true conditional independence structure underlying the data generation process by estimating the partial correlations between regions and enforcing the true 0's through the BDFR approach devised in Section \ref{sec:graph_sel}. The model is also able to recover the most likely state transition sequence for each subject, as determined by the {\it maximum a posteriori} state estimate at each time point. See Figure \ref{sim_state_mat}. It is also important to assess the ability of the method to identify true change points in the connectivity structure. Figure \ref{sim_change_point} reports the estimated connectivity change points for a representative subject.  PIBDFC is able to estimate the state sequence well while tying together the increased rate of appearance of state 3 when the stimulus changes from 0 to 1 halfway through the simulated experiment. All models were compared in terms of computation time as reported in Table \ref{sim1_table}. PIBDFC is also able to draw as many posterior draws in a third of the computation time.

\begin{table}[H]
\centering
\begin{tabular}{l|l|lll}
Metric & Method     & State 1             & State 2            & State 3            \\ \hline
\multirow{3}{*}{Edge TPR}        & PIBDFC & $0.9814$ $(0.015)$ & $1.0000$ $(0)$ & $0.9806$ $(0.010)$\\
       & Tapered SW &$0.9779$ $(0.018)$ & $0.9676$  $(0.077)$ & $0.9776$  $(0.015)$\\
       & BDFC       & $0.9221$  $(0.064)$ & $0.9435$  $(0.082)$ & $0.8326$ $(0.093)$ \\ \hline
\multirow{3}{*}{Edge TNR}        & PIBDFC &$0.9672$    $(0.007)$& $0.9585$ $(0.007)$ & $0.9351$   $(0.013)$\\
       & Tapered SW &$0.7623$ $(0.074)$  & $0.700$  $(0.107)$ & $0.7034$ $(0.104)$\\
       & BDFC       &  $0.9737$ $( 0.039)$  & $0.9835$ $(0.031)$  & $0.9822$ $(0.034)$ \\ \hline
\multirow{3}{*}{Edge F1 Score}   & PIBDFC & $0.9493$ $(0.019)$  & $0.9585$ $(0.007)$ & $0.9170$ $(0.020)$\\
       & Tapered SW &  $0.7459$ $(0.072)$  & $0.6839$  $(0.141)$ & $0.6888$ $(0.105)$\\
       & BDFC       & $0.9020$  $(0.090)$ & $0.9330$ $(0.101)$  & $0.8242$ $(0.108)$ \\ \hline
\multirow{3}{*}{State Acc}       & PIBDFC & $0.9967$ $(0.001)$  & $0.9880$ $(0.002)$ & $0.9959$  $(0.001)$\\
       & Tapered SW & $0.9340$ $(0.084)$   & $0.7496$ $(0.323)$  & $0.9538 $ $(0.113)$\\
       & BDFC       &$0.9993$ $(0.001)$  & $0.9871$  $(0.005)$ & $0.9980$ $(0.001)$ \\ \hline
\multirow{3}{*}{Comp Time (min)} & PIBDFC & \multicolumn{3}{c}{$197.57$ $(24.788)$}                       \\
       & Tapered SW & \multicolumn{3}{c}{$0.6573 $ $(0.085 )$}                        \\
       & BDFC       & \multicolumn{3}{c}{$1015.5$  $(58.922)$}                       
\end{tabular}
\caption{Simulation Study 1: results  over 30 repetitions. We report  sensitivity and specificity metrics for the estimated graphs of the corresponding states, together with the overall accuracy of the estimated state sequences. Standard deviations across the 30 simulations are showed in brackets. The proposed method maintains the best balance between sensitivity and specificity as well as latent state estimation accuracy.}
\label{sim1_table}
\end{table}

\paragraph{Simulation Study 2:}
In this second simulation study, we   measure the performance of our approach with synthetic data that are similar to real fMRI data. More specifically,  we use the Matlab simulation toolbox SimTB of \citet{Erhardt2012} and follow the simulation approach of \citet{Warnick2018b}. The SimTB toolbox implements a canonical hemodynamic response function \citep{Lindquist:2009a}, defined as a linear combination of two gamma functions, to simulate fMRI time series. This function is then convolved with a box stimulus function where Gaussian noise with variance = 0.01 is added. FC is then obtained by predefining cliques, i.e. clusters of regions, that have signal (here, 0.5) added to or subtracted from all regions in the clique simultaneously at random time points within a connectivity state. This  induces correlation while having non-Gaussian errors. We then simulate the state sequence over $T = 150$ time points with $x_t$ being 0 for the first 75 time points and 1 for the last 75 among all subjects. Similar to Simulation Study 1, we use the exact same $Q_t$ among all subjects. We repeat this process for $N = 30$ subjects over 30 simulation replicates.

\begin{table}[H]
\centering
\begin{tabular}{l|l|lll}
Metric & Method     & State 1             & State 2            & State 3            \\ \hline
\multirow{3}{*}{Edge TPR}        & PIBDFC & $1$ $(0)$          & $0.8290$  $(0.032)$ & $0.7652$ $(0.039)$ \\
       & Tapered SW & $1$ $(0)$           & $1$ $(0)$          & $1$ $(0)$          \\
       & BDFC       & $0.9769$ $(0.070) $ & $0.9014$ $(0.156)$ & $0.7203$ $(0.189)$ \\ \hline
\multirow{3}{*}{Edge TNR}        & PIBDFC & $0.9278$ $(0.004)$ & $0.8604$ $(0.041)$  & $0.9250$ $(0.040)$ \\
       & Tapered SW & $0.3286$ $(0.109)$  & $0.4583$ $(0.165)$ & $0.2500$ $(0.157)$ \\
       & BDFC       & $0.8294$ $(0.150)$  & $0.8552$ $(0.148)$ & $0.9531$ $(0.088)$ \\ \hline
\multirow{3}{*}{Edge F1 Score}   & PIBDFC & $0.9278$ $(0.004)$ & $0.7134$ $(0.045)$  & $0.7083$ $(0.055)$ \\
       & Tapered SW & $0.3286$ $(0.109)$  & $0.4583$ $(0.165)$ & $0.2500$ $(0.157)$ \\
       & BDFC       & $0.8063$ $(0.138)$  & $0.7717$ $(0.192)$ & $0.6822$ $(0.176)$ \\ \hline
\multirow{3}{*}{State Acc}       & PIBDFC & $0.8526$ $(0.029)$ & $0.7507$ $(0.022)$  & $0.7727$ $(0.022)$ \\
       & Tapered SW & $0.7199$ $(0.175)$  & $0.4133$ $(0.100)$ & $0.6342$ $(0.108)$ \\
       & BDFC       & $0.6110$ $(0.43)$   & $0.7181$ $(0.11)$  & $0.5541$ $(0.37)$  \\ \hline
\multirow{3}{*}{Comp Time (min)} & PIBDFC & \multicolumn{3}{c}{$161.23$ $(29.493)$}                       \\
       & Tapered SW & \multicolumn{3}{c}{$1.9241 $ $(0.31)$}                        \\
       & BDFC       & \multicolumn{3}{c}{$500.57$ $(18.11)$}                       
\end{tabular}
\caption{Simulation Study 2: results  over 30 repetitions. We report  sensitivity and specificity metrics for the estimated graphs of the corresponding states, together with the overall accuracy of the estimated state sequences. Standard deviations across the 30 simulations are shown in brackets. The proposed method maintains the best balance between sensitivity and specificity as well as latent state estimation accuracy.}
\label{sim2_table}
\end{table}

In Table \ref{sim2_table} we show the results to the application on the SimTB data. PIBDFC does a good job at detecting the connectivities  between the simulated regions, despite a misspecified likelihood. The performance in both graph and state estimation  appears to decline slightly in comparison to the Simulation 1 setting, which is expected. The Tapered SW approach suffers from low specificity. Compared to the standard HMM of BDFC, the proposed PIBDFC performs slightly better at detecting changes in state transitions, thus improving graph estimation performance as a result. This is likely due to the distortion introduced in the partial correlation by the convolution with the hemodynamic response function. In this setting, the covariate information becomes more relevant in helping the model identify changes in the state transition behavior. The computational time is also quite favorable compared to the approach of \citet{Warnick2018b}, despite allowing for individual differences in state dynamics among the 30 subjects.

\paragraph{Simulation Study 3:}

In this simulation setting, we compare the performances of our model and the Connectivity Change Point Detection (CCPD) model of \citet{Kundu2018} on edge- and change-point detection. Contrary to our model, the CCPD model employs a two-stage approach for estimating dynamic FC. In the first stage, the model learns the number and locations of the change points from all available subjects' data. In the second stage, a graphical lasso approach is applied independently to the time scans between two change points. Since the CCPD model assumes that every change point occurs at the same time for each subject, in order to fairly compare the two methods we simulate data under the CCPD assumption of common change points. More specifically, we set $T=300$ and generate $Y^i_t \sim N(0,\Omega_{s_t})$ where $s_t$ varies across the following sequence of states:\{1, 2, 3, 1\} switching at $t= 75, 150, 225$, for a total of 3 change-points overall. We use the same true partial correlation matrices to generate the data as in Simulation study 1.  For the PIBDFC, a time point $t$ for subject $i$ was judged to be a change point if $P(s_t^i \neq s_{t-1}^i|Y^i_{1:T}) > 0.95$. PIBDFC does not assume common change points and, as a result, does not infer common change points across individuals; therefore, we report the average  number of change points across all subjects.

\begin{table}[H]
\centering
\resizebox{\columnwidth}{!}{
\begin{tabular}{llll|lll}
\multicolumn{1}{l|}{Method}   & \multicolumn{3}{c|}{PIBDFC}  & \multicolumn{3}{c}{CCPD} \\ \hline
\multicolumn{1}{l|}{State}             & 1        & 2       & 3      & 1        & 2        & 3        \\ \hline
\multicolumn{1}{l|}{Edge TPR} & $0.9650 $ $(0.02)$ & $1.0000$  $(0)$& $0.9867$ $(0.01)$ & $0.9333$ $(0.02)$  & $1.0000$   & $0.9800$ $(0.02)$   \\
\multicolumn{1}{l|}{Edge TNR} & $0.9674$ $(0.01)$  & $0.9719$ $(0.01)$ & $0.9615$ $(0.02)$& $0.9733$ $(0.09)$  & $0.9978$  $(0.01)$& $0.7719$  $(0.06)$  \\ 
\multicolumn{1}{l|}{Edge F1 Score} & $0.9336$ $(0.02) $  & $0.9719$ $(0.01) $ & $0.9486$ $ (0.02)$ & $0.9078$ $(0.08)$  & $0.9978$ $(0.01)$  & $0.7564$ $(0.06)$  \\ \hline
\multicolumn{1}{l|}{Num ChgPts (3)}    & \multicolumn{3}{c|}{$3.8$ $(0.97)$} & \multicolumn{3}{c}{$3.1$ $(0.38)$}  

\end{tabular}
}
\caption{Simulation Study 3: Results over 30 repetitions. We show the entry-wise true positive and true negative rates for the estimated graphs for the corresponding states. We also show the estimated number of  change points. PIBDFC performs comparably to CCPD in the setting where change points are common among subjects despite no explicit assumption of this being the case. }
\label{sim3_table}
\end{table}

In Table \ref{sim3_table}, we show the results of the comparison between PIBDFC and CCPD under a shared change point model. CCPD is indeed able to accurately detect the number of change points and the resulting graph structure in each partition well. By thresholding the posterior probability of a change point, our model tends to overestimate the number of change points on average, as it sometimes estimates very sudden changes of state for a brief collection of time points in some subjects. In contrast, in simulation studies 1 and 2, the change points are generated from a process that truly follows a hidden Markov model, leading to more accurate estimates. By leveraging on the assumption of common change points, the two-stage CCPD model can achieve increased accuracy, while our model allows for the incorporation of individual transitions and covariates in the transition probabilities.

\section{Case Study}

We apply the proposed PIBDFC model to the motivating dataset. In our application, we demonstrate how the model can recover expected change points in dynamic FC states, as those states align quite well with the experimental events regulated by the behavioral task. We are also able to estimate the effect of pupil dilation on the subjects' propensity to change states.

\subsection{Experimental design and data collection}

In this experiment, subjects performed a handgrip task adapted from \citet{mather_isometric_2020}.  Thirty-one participants (18 females, mean age 25 years $\pm$ 4 years) enrolled in this study at the University of California, Riverside Center for Advanced Neuroimaging, but one was excluded due to a history of attention deficit hyperactive disorder resulting in a total of $N = 30$ subjects. All subjects provided written informed consent to participate, and received monetary compensation for their participation. The study protocol was approved by the University of California, Riverside Institutional Review Board (IRB). Magnetic resonance imaging (MRI) data were collected on a Siemens 3T Prisma MRI scanner (Prisma, Siemens Healthineers, Malvern, PA) with a 64 channel receive-only head coil. fMRI data were collected using a 2D echo planar imaging sequence (echo time (TE) = 32 ms, repetition time (TR) = 2000 ms, flip angle = $77^\circ$, and voxel size = $2 \times 2 \times 3 \, mm^3$ , slices=52) while pupillometry data were collected concurrently with a TrackPixx system (VPixx, Montreal, Canada). 

All subjects underwent a 12.8-minute experiment in which they alternated between six resting state blocks and five squeeze blocks. In the squeeze blocks, they brought their dominant hand to their chest while holding a squeeze-ball  \citep{mather_isometric_2020}.  The five squeeze blocks lasted 18 seconds while the interspersed six resting state blocks had durations of five-, two-, two-, five-, one-, and one-minute, respectively.

All subjects underwent two sessions: one where they executed the squeeze at maximum grip strength (active session), and one where they still brought their arm up to their chest but were instructed simply to touch the ball and not to squeeze it (sham session).  The fMRI data underwent a standard preprocessing pipeline using the brain software library (FSL). The pipeline consisted of slice time correction, motion correction, susceptibility distortion correction, and spatial smoothing using a kernel Gaussian smoothing factor set at a full-width half maximum of 0.8475 \citep{smith_advances_2004, woolrich_bayesian_2009}. Finally, all data were transformed from the individual subject space to the Montreal Neurological Institute (MNI) standard space using standard procedure in FSL \citep{smith_advances_2004, woolrich_bayesian_2009}.

Pupillometry data were collected during the scans, using a sampling rate of 2kHz, preprocessed using the ET-remove artifacts toolbox (github.com/EmotionCognitionLab/ET-remove-artifacts), and downsampled to match the temporal resolution of the fMRI data \citep{mather_isometric_2020}. To measure pupil dilations relative to baseline, the dataset was normalized by dividing by subject-specific means of the first five-minute resting state block (prior to any squeeze or hand-raising), leading to percent signal changes. Three subjects' data were discarded due to technical difficulties during the acquisition of pupil dilation measurements, resulting in $N = 27$ for all subsequent analyses.

Since we used a pseudo-resting state paradigm, 
our interest was focused on five networks of interest that have all been associated with resting state and have been related to attention in some manner. 
Default mode network (DMN; a resting state network) and dorsal attention network (DAN; an attention network) were selected because squeezing ought to invoke a transition from the resting state into a task-positive state \citep{greicius_default-mode_2004}.  The fronto-parietal control network (FPCN) was chosen because it is linked to DAN and regulates perceptual attention (Dixon et al., 2018). Salience network (SN) was selected because it determines which stimuli in our environment are most deserving of attention \citep{mather_isometric_2020, menon_saliency_2010}. Talariach coordinates for regions of interest (ROIs) within DMN, FPCN, and DAN were taken from \citet{deshpande_instantaneous_2011} and converted to MNI coordinates while SN MNI coordinates were taken directly from \citet{raichle_restless_2011} \citep{deshpande_instantaneous_2011, laird_brainmap_2005, lancaster_bias_2007, raichle_restless_2011}.  Two ROIs from FPCN (dorsal anterior cingulate cortex and left dorsolateral prefrontal cortex) were excluded due to their close location to other ROIs.  The locus coeruleus (LC) was localized using the probabilistic atlas described in \citet{langley_characterization_2020}.  Blood oxygen level-dependent (BOLD) signal from each voxel within an ROI were extracted and averaged to represent the overall signal for an ROI. 
We eventually considered 31 total ROIs: 9 from DMN, 7 from FPCN, 6 from DAN, 7 from SN, and 2 from LC. The MNI anatomical coordinates for the four attention networks and LC were used to center a 5 mm$^3$ isotopic sphere \citep{deshpande_multivariate_2009, stilla_posteromedial_2007}. See the Appendix for a list of the ROIs and corresponding MNI stereotaxic space coordinates and networks.

\subsection{Model fitting}
The 31 ROIs described above formed the vectors of BOLD responses $Y^i_t = (Y^i_{t1},\ldots, Y^i_{t31})$ measured on subject $i=1, \ldots, 27$ at time t, for $t = 1,\ldots, 1050$. We also included concurrently recorded pupillometry data as a proxy for quantifying the effect of LC engagement on the dynamics of FC  \citep{joshi_context-dependent_2022}.

We fit our model with different number of total states, i.e., $S = 3, 4, 5, 6$. However, when assuming more than 3 states, the fit simply degenerated to 3 states in the posterior inference, with no observations assigned to additional states. This result indicates no posterior support for models with $S>3$ Thus,  here we present the model specification for 3 states with the following settings for the hyperparmeters in  \eqref{eq:trans_model}. We set the group level baseline relative transition prior means $z^0_{rr} = 2$ for $r = 2,3$ while all other elements of $z^0_{\cdot \cdot}$ are set to 0. We also set the prior spread of the baseline transitions and pupillary effects  $\sigma_z, \sigma_\eta = 0.05$. This combination of settings is used to encourage self-transitions, as they correspond to preferring smoother state sequences a priori among all subjects. We set the prior variability of the subject-level transition parameters around the group-level transition parameters, by choosing $\sigma_\xi, \sigma_\rho = 0.1$, therefore capturing individual differences between subjects on the log-odds of transitioning between states. Lastly, $\tau_0$, the hyperparameter informing prior knowledge of connectivity network sparsity, is set to 1, as this value corresponds to a prior distribution with a high spread over edge densities (see Figure \ref{tau_0_effects} in the Appendix). 

\begin{figure}[t!]
    \centering
    \includegraphics[width = 0.95\textwidth, clip]{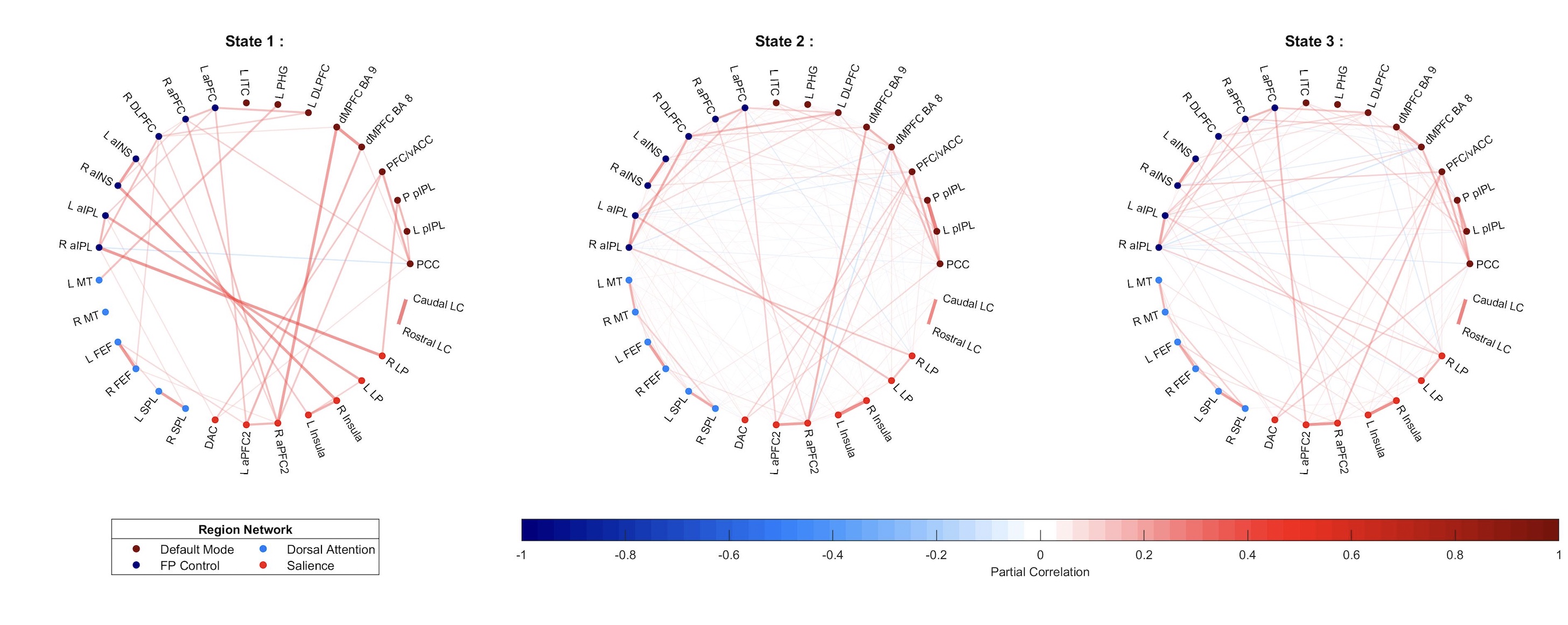}
    \caption{Real Data Analysis: the estimated connectivity networks for the ROIs. Nodes represent ROIs and the edges denote the partial correlations between the connected nodes. The edge colors correspond to the directionality of the partial correlations and the width corresponds to the magnitude. Node colors identify clusters of regions into \emph{a priori} defined networks. See Section \ref{sec:realdata} and Table \ref{ROI_desc} in the Appendix} 
    \label{conn_graph}
\end{figure}

\subsection{Results and Inference}
\label{sec:realdata}

\begin{figure}[t!]
    \centering
    \includegraphics[width = 0.95\textwidth]
    {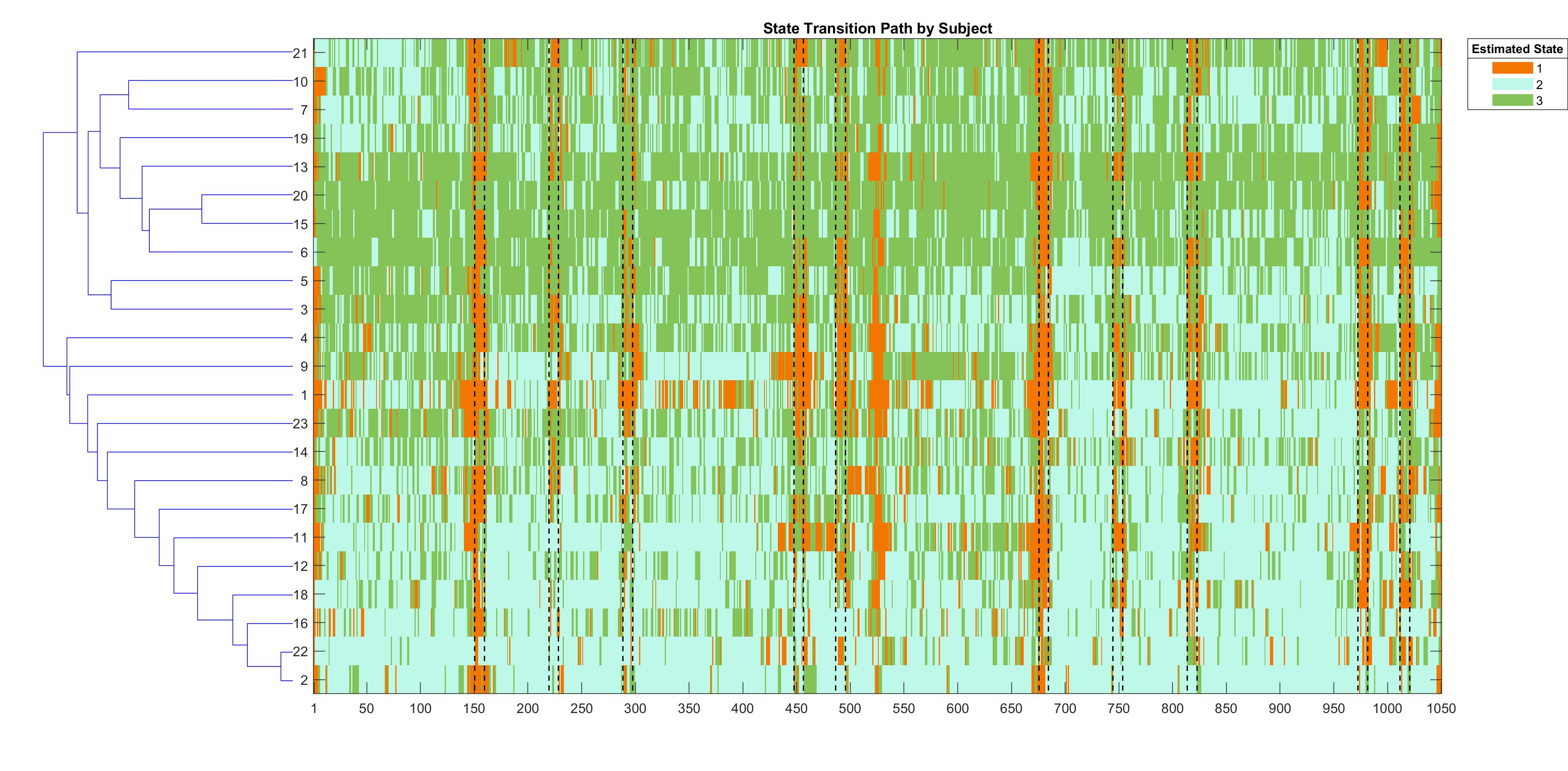}
    \caption{Real Data Analysis: Estimated states' transition path for each subject. The horizontal axis indicates the TR with vertical dotted lines indicating portions where the subject raises their arm. Subject sequences are aligned so that the first 525 time points show sequences from the sham condition and the time points 526-1050 show sequences from the active condition. The vertical axis displays the subject indices, ordered by similarity in state trajectory according to a hierarchical clustering (based on the Euclidean distance) of their MAP transition behavior.  }
    \label{state_seq}
\end{figure}

Figure \ref{conn_graph} plots the estimated connectivity networks for each of the three states. Nodes represent ROIs and edges identify the estimated non-zero partial correlations between pairs of nodes. The edge colors correspond to the directionality of the partial correlations and the width corresponds to the magnitude. The dotted colors in the nodes identify clusters of regions within \emph{a priori}, knowledge-based, neuroscientific networks (from the top right section in counter-clockwise order): Default Mode Network (DMN), Frontal Parietal Control Network (FPCN), Dorsal Attention Network (DAN), Salience Network (SN), and Locus Coeruleus (LC). 
Figure \ref{state_seq} shows the \emph{maximum a posteriori} (MAP) estimated state sequences from our model for all 27 subjects. The subjects' rows are ordered by the similarity of the estimated state trajectories as captured by a hierarchical clustering using Euclidean distance. 


By inspecting Figure \ref{conn_graph}, it is apparent that state 1 shows relatively sparser connectivity than the other two states. In state 1, we can see strong bilateral connectivity among homologous regions in the left and right hemispheres, as well as several nodes in FPCN (dark blue) showing strong connectivity with multiple nodes in SN (light red); likewise, several nodes in DMN (dark red) show connectivity with SN (light red) nodes. There is almost no presence of anti-correlation. The dominance of SN connectivities together with both DMN and FPCN suggests that arousal may be up-regulated in this state. Indeed,   Figure \ref{state_seq} suggests that state 1 occurs predominantly during the 'squeeze' periods of the behavioral task, when subjects either squeezed the squeeze ball or held it to their chest. This observation suggests that our model was able to detect those objectively-definable events in the time series of this experimental dataset. 


In state 2, we see a quite different pattern: weaker average connectivity when compared to state 1, but also many more of these weaker connections both within-network and between networks. In addition to relatively ubiquitous within-network connections within FPCN (dark blue) and DMN (dark red), state 2 is characterized by cross-network connectivity -- and anti-connectivity -- between DMN and FPCN. Interestingly, these parallel some of the strongest connectivities from state 1. The relative occupancy in state 2 appears higher in the active condition (Figure \ref{state_seq}, right half) than the sham condition (Figure \ref{state_seq}, left half), suggesting subjects tended to occupy this relatively strong, broadly-connected state more often when periodically engaging in actively squeezing the ball. 

The strongest connections in state 3 deviate from those identified in states 1 and 2. There is weaker overall connectivity than state 1, but the connections are stronger and sparser (fewer connections) than state 2. We do again see many within-network connections, as well as relatively strong connections between nodes in FPCN (dark blue) and SN (light red), and also again between DMN (dark red) and SN (light red). However, we also see many more connections with SN from DAN (light blue) than in either of the other two states. We can therefore characterize this state as more sparsely connected than state 2 but still with broad connectivity, which is also consistent with the differences visually apparent in this state between active and sham conditions (right and left halves of Figure \ref{state_seq}): this state traded off with state 2 for relative percentage occupancy across the subjects.

\begin{figure}[t!]
    \centering
    \includegraphics[width = 0.95\textwidth]{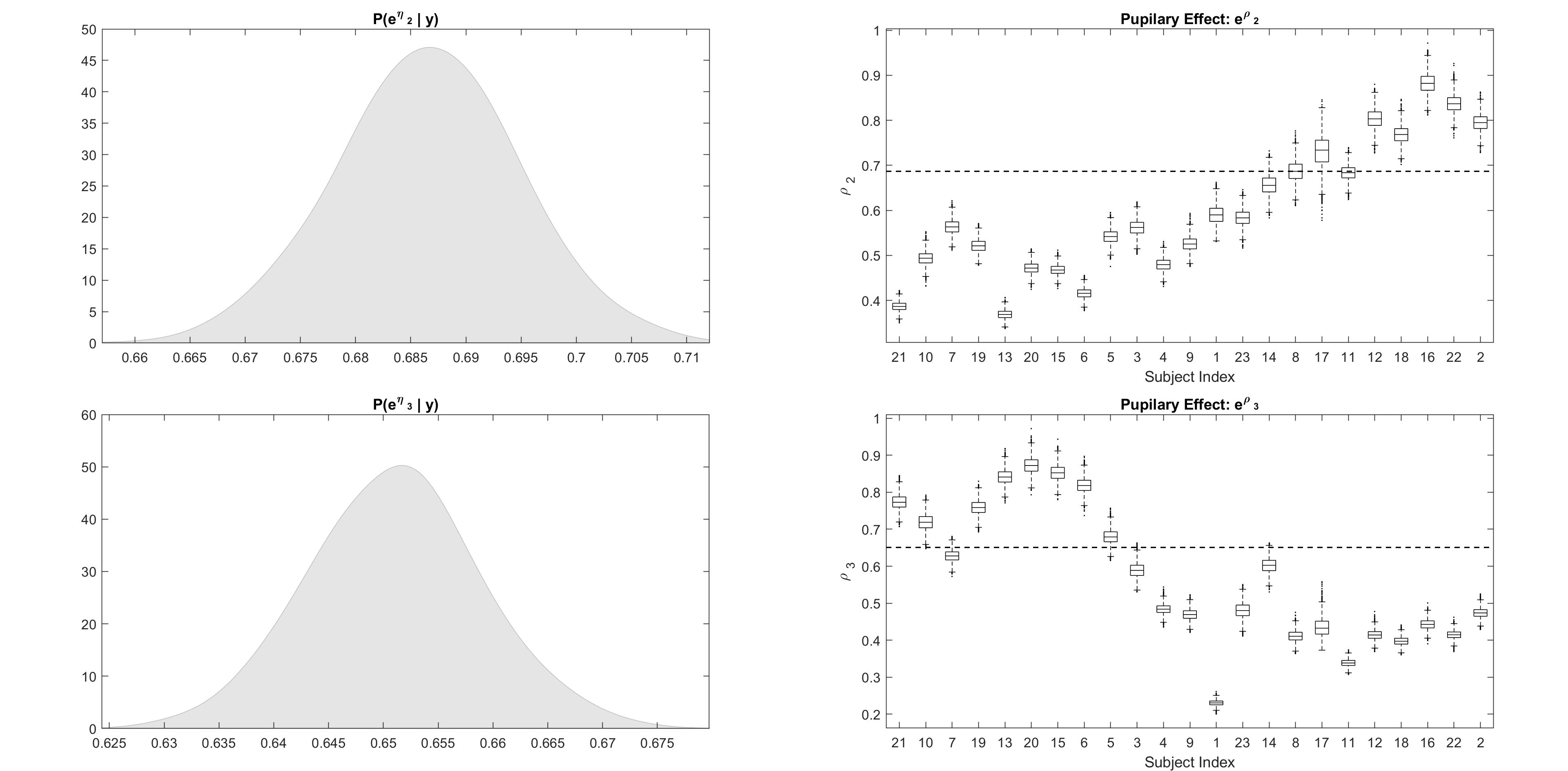}
    \caption{Real Data Analysis: The posterior distribution of the group effect of pupillary dilation $e^\eta$ (left), and individual effects of pupillary dilation $e^\rho$. Rows indicate the propensity for transitioning into states 2 and 3 respectively. For the individual effects, subjects are identically ordered as in Figure \ref{state_seq}. The horizontal dotted line is the posterior mean for the group-level effects, $\eta_2 = 0.687$ and $\eta_3 = 0.651$ respectively.}
    \label{rho_eta}
\end{figure}

Finally, a unique feature of our model is that it allows the investigation of how pupillary dilation modulates state transitions. Figure \ref{rho_eta} provides the posterior distribution of the group ($e^\eta$, left) and individual ($e^\rho$) effects of pupil dilation on state dynamics. We start by assessing the relationship between pupil dilation and state transitions for the group. Based on our findings, a $1\%$ increase in pupil dilation relative to baseline is associated with a $31.4\%$ ($95\% CI: 29.7\% - 32.9\%$) decrease in the odds of transitioning to state 2 and a  $34.9\%$ ($95\% CI: 33.3\% - 36.4\%$) decrease in the odds of transitioning to state 3, in comparison to remaining in the baseline state (state 1). This result is coherent with the findings outlined above since increased pupil dilation (a proxy for increased arousal/effort) appears associated with transitioning toward the less densely connected connectivity structure of state 1, dominated by edges between SN and both DMN and FPCN. We should note that the causal direction of the inferred associations 
can not be investigated by this model. 

Further inspection of the right column of Figure \ref{rho_eta} shows that the posterior distributions of the individual effects of pupil dilation $e^{\rho_{\cdot \cdot}}$ is decidedly below 1 for all subjects, i.e.  the association between increased pupil dilation and state 1 holds for all subjects measured. Subjects are ordered along the horizontal axis according to their similarity in state trajectories obtained from a hierarchical clustering, based on the Euclidean distance (similarly as in Figure \ref{state_seq}). The horizontal dashed line represents the posterior mean from the group estimate in the right panel. It is interesting to note the differing clusters when comparing the posterior distributions of $e^{\rho_{2 \cdot}}$ to $e^{\rho_{2 \cdot}}$: trending downwards and upwards respectively. Quite importantly, the correspondence between the groupings observed in Figure \ref{state_seq} and  Figure \ref{rho_eta} is a result of the posterior inference, not necessarily implied by the structure of our model. The  differences in state trajectories between subjects lie in the state occupancy when pupil dilation is not higher than the reference, despite all subjects tending to transition to state 1 when raising their arm. 

More specifically, subjects clustered in the first half of Figure \ref{rho_eta} (right) tend to occupy state 3 during non-squeeze sections and so are even more likely to transition away from state 2 during periods of high pupil dilation. Similarly,  subjects in the second half of the Figure tend to occupy state 2 during non-squeeze sections, and are thus very likely to transition away from state 3. This heterogeneity is important as it provides a more thorough understanding of the relationship between increased pupil dilation and transitions toward different cognitive states. 


\section{Discussion}
 We have proposed a multi-subject Bayesian approach for estimating dynamic FC where the brain network state transitions are dynamically informed by concurrently-recorded subject-specific covariates. The proposed method allows for group-level and subject-level inferences on the effects of time-varying covariates on the connectivity dynamics. We have applied our model to multi-subject resting state fMRI data with pupillary physiological data and we have shown associations between pupil dilation and strengthened connectivity between the SN brain regions with both the FPCN and DMN. This association coinciding with subject arm-raising/squeezing suggests SN connections with both FPCN and DMN are associated with subject arousal. 
 
While we focused here on covariates that were concurrently recorded on each  subject, our model can also incorporate covariates that are subject-specific and not time-varying. For example, demographic information may be added to the regression terms in \eqref{eq:trans_model}--\eqref{eq:logrelat} and inform subject-specific transition probabilities to describe individual variability over the entire fMRI experiment.
 
 Our model assumes a maximum number of states $S$ to be pre-specified \textit{a priori}. In our application, only a subset of the $S$ available states was visited. However, in general, the number of states could be learned by assuming a Bayesian-nonparametric specification where the number of FC states is learned directly from the data \citep[see, for example,][]{Beal2002,fox2011}. However, the computational complexity of the inferential algorithm would increase considerably. Variational Bayes approaches could be implemented to obtain approximate inferences on the network connections.  
 
 Finally, the individual connectivity patterns could be associated with clinical or behavioral outcomes, e.g., to examine the individual heterogeneity of responses to treatments. A two-stage scalar-on-image approach can be devised where the posterior means of the precision matrices are obtained from our model in the first stage and then used as predictors to investigate the association with the outcome in the second stage. These directions of research will be the object of future investigations.
  

\subsection*{Acknowledgements}
Jaylen Lee has been supported by the National Science Foundation Graduate Research Fellowship Grant No. DGE-1839285. 

\subsection*{Code}
 The code for the proposed PIBDFC model can be downloaded for the following GitHub repository:\\ \url{https://github.com/jayesrule/PIBDFC}.

\newpage

\section*{Appendix}

\subsubsection*{Appendix 1}
The following table reports the list of ROIs employed in the case study along with corresponding  MNI stereotaxic space coordinates and their classification in  \emph{a priori}  defined networks. 

\begin{table}[H]
\resizebox{\columnwidth}{!}{%
\begin{tabular}{llll}
Network                                          & Abbreviation & Full Name                                              & MNI Coordinates     \\ \hline
\multirow{9}{*}{\rotatebox[origin=c]{90}{\parbox[c]{2cm}{\centering Default Mode Network}}}            & PCC          & Posterior Cingulate Cortex                             & (2, 54, 16)         \\
                                                 & L pIPL       & Left Posterior Inferior Parietal Lobule                & (-46, -72, 28)      \\
                                                 & R pIPL       & Right Posterior Inferior Parietal Lobule               & (50, -64, 26)       \\
                                                 & PFC/vACC     & Orbitofrontal Cortex/Ventral Anterior Cingulate Cortex & (4, 30, 26)         \\
                                                 & dMPFC BA 8   & Dorsomedial Prefrontal Cortex Broadmann Area 8         & (-14, 54, 34)       \\
                                                 & dMPFC BA 9   & Dosomedial Prefrontal Cortex Brodmann Area 9           & (22, 58, 26)        \\
                                                 & L DLPFC      & Dorsolateral Prefrotnal Cortex                         & (-50, 20, 34)       \\
                                                 & L PHG        & Parahippocampal Gyrus                                  & (-10, -38, -2)      \\
                                                 & L ITC        & Inferolateral Temporal Cortex                          & (-60, -20, -18)     \\ \hline
\multirow{9}{*}{\rotatebox[origin=c]{90}{\parbox[c]{2cm}{\centering Fronto-Parietal Control Network}}}  & L aPFC       & Left Anterior Prefrontal Cortex                        & (-36, 56, 10)       \\
                                                 & R aPFC       & Right Anterior Prefrontal Cortex                       & (34, 52, 10)        \\
                                                 & dACC         & Dorsal Anterior Cingulate Cortex                       & N/A                 \\
                                                 & L DLPFC      & Left Dorsolateral Prefrontal Cortex                    & N/A                 \\
                                                 & R DLPFC      & Right Dorsolateral Prefrontal Cortex                   & (46, 14, 42)        \\
                                                 & L aINS       & Left Anterior Insula                                   & (-30, 20, -2)       \\
                                                 & R aINS       & Right Anterior Insula                                  & (32, 22, -2)        \\
                                                 & L aIPL       & Left Anterior Inferior parietal Lobule                 & (-52, -50, 46)      \\
                                                 & R aIPL       & Right Anterior Inferior Parietal Lobule                & (52, -46, 46)       \\ \hline
\multirow{6}{*}{\rotatebox[origin=c]{90}{\parbox[c]{2cm}{\centering Dorsal Attention Network}}}      & L MT         & Left MidThalamus                                       & (-44, -64, -2)      \\
                                                 & R MT         & Right MidThalamus                                      & (50, -70, -4)       \\
                                                 & L FEF        & Left Frontal Eye Field                                 & (-24, -8, 50)       \\
                                                 & R FEF        & Right Frontal Eye Field                                & (28, -10, 50)       \\
                                                 & L SPL        & Left Superior Parietal Lobule                          & (-26, -52, 56)      \\
                                                 & R SPL        & Right Superior Parietal Lobule                         & (24, -56, 54)       \\ \hline
\multirow{7}{*}{\rotatebox[origin=c]{90}{\parbox[c]{2cm}{\centering Salience Network}}}               & DAC          & Dorsal Anterior Cingulate                              & (0, -22, 36)        \\
                                                 & L aPFC       & Left Anterior PFC                                      & (-34, 44, 30)       \\
                                                 & R aPFC       & Right Anterior PFC                                     & (32, 44, 30)        \\
                                                 & L Insula     & Left Insula                                            & (-40, 2, 6)         \\
                                                 & R Insula     & Right Insula                                           & (42, 2, 6)          \\
                                                 & L LP         & Left Lateral Parietal                                  & (-62, -46, 30)      \\
                                                 & R LP         & Right Lateral Parietal                                 & (62, -46, 30)       \\ \hline
\multirow{2}{*}{Locus Coeruleus}                 & R LC         & Rostral Locus Coeruleus                                & Probabilistic Atlas \\
                                                 & C LC         & Caudal Locus Coeruleus                                 & Probabilistic Atlas
\end{tabular}
}
\caption{The ROIs used in the case study along with \emph{apriori}  defined networks. }
\label{ROI_desc}
\end{table}

\subsection*{Appendix 2}

\begin{figure}[t]
    \centering
    \includegraphics[width=0.8\textwidth, trim = 4cm 8cm 4cm 8cm, clip]{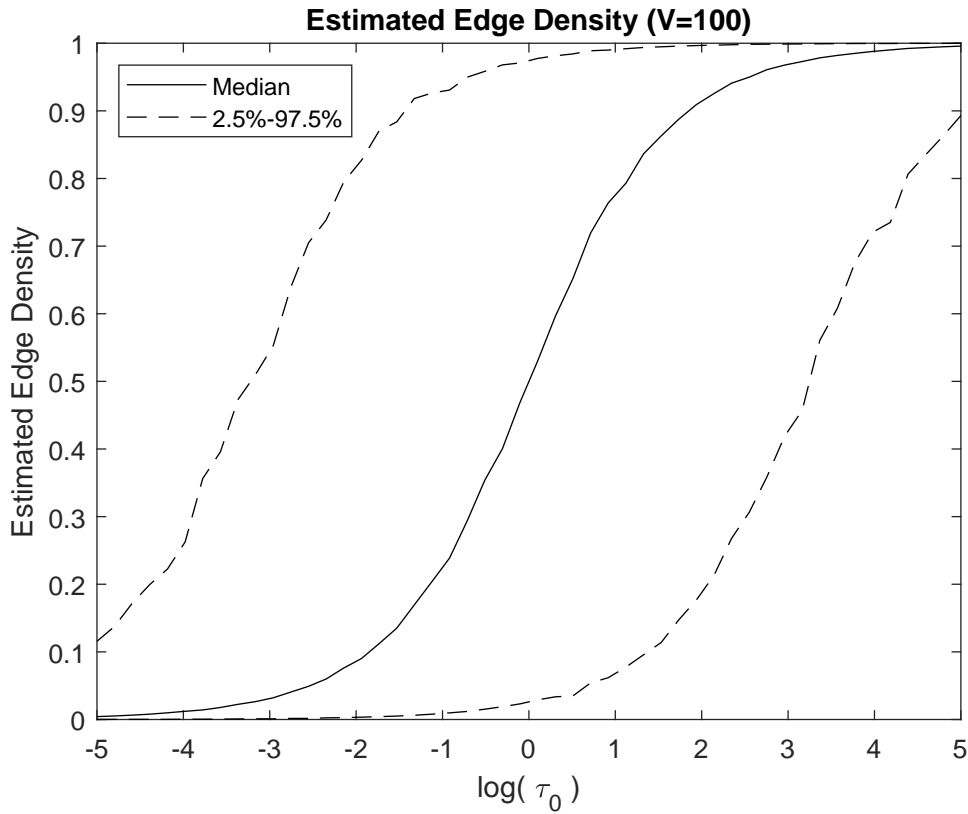}
   \caption{For each value of $\tau_0$, we simulate 1000 undirected $100\times 100$ graphs under the graphical horseshoe prior. 
   Plotted above are the 2.5, 50, and 97.5 percentiles of the edge density as a function of $\tau_0$. A value of $\tau_0=1$ leads to approximately a 50\% expected edge density, with high spread, in the sampled graphs.  }
    \label{tau_0_effects}
\end{figure}

Figure \ref{tau_0_effects} illustrates how to to specify the value of the parameter $\tau_0$, by simulating 1,000 undirected graphs from the model. A larger $\tau_0$ is associated with higher expected edge densities \textit{a priori}. Additionally, we find that a $\tau_0 = 1$ gives an expected edge density of approximately 50\% while having the largest spread.

\newpage
\bibliographystyle{plainnat}
\bibliography{references.bib}

\end{document}